\documentclass[12pt]{article}

\usepackage{graphics,graphicx}
\usepackage{amsmath}
\usepackage{times}
\usepackage{amstext}
\usepackage{amssymb}
\usepackage{bbm}

\def\Det{\mbox{Det}}
\def\Tr{\mbox{Tr}}
\def\rank{\mbox{rank}}
\def\beq{\begin{equation}}
\def\eeq{\end{equation}}
\def\bea{\begin{eqnarray}}
\def\eea{\end{eqnarray}}
\providecommand{\openone}{\leavevmode\hbox{\small1\kern-3.8pt\normalsize1}}
\newtheorem{theorem}{Theorem}
\newtheorem{prop}{Proposition}
\newtheorem{corollary}{Corollary}
\newtheorem{lemma}{Lemma}

\newcommand{\rr}{\mathbbm{R}}

\begin{document}

\title{Multi-mode bosonic Gaussian channels}
\author{F.\ Caruso$^{1}$, J.\ Eisert$^{2}$, V.\ Giovannetti$^{1}$, and
A.S.\ Holevo$^{3}$  \\
{\small $^{1}$ NEST CNR-INFM \& Scuola Normale Superiore,}
{\small Piazza dei Cavalieri 7, I-56126 Pisa, Italy}
\\ {\small $^{2}$ Institute for Mathematical Sciences, Imperial
College London,} {\small London SW7 2PE, UK}
\\
{\small $^{3}$Steklov Mathematical Institute,} {\small Gubkina 8,
119991 Moscow, Russia}
 }
\maketitle

\begin{abstract}
A complete analysis of multi-mode bosonic Gaussian channels is
proposed. We clarify the structure of unitary dilations of general
Gaussian channels involving any number of bosonic modes and present
a normal form. The maximum number of auxiliary modes that is needed
is identified, including all rank deficient cases, and the specific
role of additive classical noise is highlighted. By using this
analysis, we derive a canonical matrix form of the noisy evolution
of $n$-mode bosonic Gaussian channels and of their weak
complementary counterparts, based on a recent generalization of the
normal mode decomposition for non-symmetric or locality constrained
situations. It allows us to simplify the weak-degradability
classification. Moreover, we investigate the structure of some
singular multi-mode channels, like the additive classical noise
channel that can be used to decompose a noisy channel in terms of a
less noisy one in order to find new sets of maps with zero quantum
capacity. Finally, the two-mode case is analyzed in detail. By
exploiting the composition rules of two-mode maps and the fact that
anti-degradable channels cannot be used to transfer quantum
information, we identify sets of two-mode bosonic channels with zero
capacity.
\end{abstract}

\tableofcontents

\
\newline
\newline
\

Bosonic Gaussian channels are ubiquitous in physics. They arise
whenever a harmonic system interacts linearly with a number of
bosonic modes which are inaccessible in principle or in practice
\cite{HOLEVOBOOK,HolevoOld,HW,EW,HolevoOld2,Demoen,FOP}. They
provide realistic noise models for a variety of quantum optical
and solid state systems when treated as open quantum systems,
including models for wave guides and quantum condensates. They
play a fundamental role in characterizing the efficiency of a
variety of tasks in continuous-variables quantum information
processing~\cite{BL}, including quantum communication~\cite{CVDM}
and cryptography~\cite{CRYPTO}. Most importantly, communication
channels such as optical fibers can to a good approximation be
described by Gaussian quantum channels.

Not very surprisingly in the light of the central status of such
quantum channels, a lot of effort has been recently devoted to
studying their properties (see Ref.~\cite{EW} for a review), based
on a long tradition of work on Gaussian channels
\cite{Demoen,HolevoOld,HW}. Specifically, from a quantum information
perspective, a key question is whether or not a channels allows for
the reliable transmission of classical or quantum information
\cite{HW,EW,Lloyd,Lloyd2,SEW,HOLEVONEW,WGG,CG,CGH,Lindblad}.
Significant progress has been made in this respect in recent years,
although for some important cases, like the thermal noise channel
modelling a realistic fiber with offset noise, the quantum capacity
is still not yet known. In this context, the degradability
properties represent a powerful tool to simplify the quantum
capacity issue of such Gaussian channels. Indeed, in
Refs.~\cite{CG,CGH} it has been shown
that with some (important)
exceptions, Gaussian channels which operate
on a single bosonic mode
(i.e., one-mode Gaussian channels) can be classified as weakly
degradable or anti-degradable. This paved the way for the
solution of the quantum capacity~\cite{QUANTUMCAP} for a large class
of these maps~\cite{WGG}.

Here, first we propose a general construction of unitary dilations
of multi-mode quantum channels, including all rank-deficient cases.
We characterize the minimal noise maps involving only true quantum
noise. Then, by using a generalized normal mode decomposition
recently introduced in Ref.\ \cite{WOLFLAST}, we generalize the
results of Refs.\ \cite{CG,CGH} concerning Gaussian weak
complementary channels to the multi-mode case giving a simple
weak-degradability/anti-degradability condition for such channels.
 The
paper ends with a detailed analysis of the two-mode case. This is
important since any $n$-mode channel can always be reduced to
single-mode and two-mode components~\cite{WOLFLAST}. We detalize the
degradability analysis and investigate a useful decomposition of a
channel with the additive classical noise map that allows us to find
new sets of channels with zero quantum capacity.

\section{Multi-mode bosonic Gaussian channels}

Gaussian channels arise from linear dynamics of open bosonic
system interacting with a Gaussian environment via quadratic
Hamiltonians. Loosely speaking, they can be characterized as CPT
maps that transform Gaussian states into Gaussian states
\cite{Survey,AI}.

\subsection{Notation and  preliminaries}

Consider a system composed by $n$ bosonic modes having canonical
coordinates $\hat{Q}_1,\hat{P}_1, \cdots, \hat{Q}_n, \hat{P}_n$. The
canonical commutation relations of the canonical coordinates,
$[\hat{R}_j, \hat{R}_{j^\prime}] = i (\sigma_{2n} )_{j,j^\prime}$,
where $\hat{R} := ( \hat{Q}_1, \cdots, \hat{Q}_n; \hat{P}_1, \cdots,
\hat{P}_n)$, are grasped by the $2n\times 2n$ {\it commutation
matrix}
\begin{equation}
    \sigma_{2n}=
    \left[
    \begin{array}{cc}
    {0} & \openone_n\\
    -\openone_n & {0}
    \end{array}
    \right] \;,
     \label{NSYMP}
\end{equation}
when this order of canonical coordinates is chosen, (here
$\openone_n$ is the $n\times n$ identity
matrix)~\cite{HW,EW,Survey}. Even though different reordering of the
elements of $\hat{R}$  will not affect the definitions that follow,
we find it useful to assume a specific form for $\sigma_{2n}$. One
defines the group of real {\em symplectic matrices} $Sp(2n,\rr)$ as
the set of  $2n\times 2n$ real matrices $S$ which satisfy the
condition
\begin{eqnarray}
    S \sigma_{2n} S^T = \sigma_{2n} \label{NSYMPD}\,.
\end{eqnarray}
Since $\Det [\sigma_{2n} ]= 1$, and $\sigma_{2n}^{-1} = -
\sigma_{2n}$, any symplectic matrix $S$ has $\Det[S]=1$ and it is
invertible with $S^{-1} \in Sp ( 2n, \rr)$. Similarly, one has $S^T
\in   Sp ( 2n, \rr)$. Symplectic matrices play a key role in the
characterization of bosonic systems. Indeed, define the {\it Weyl
(displacement) operators} as $\hat{V}(z) = \hat{V}^\dag(-z) := \exp
[ i \hat{R}  z]$ with $z:= ( x_1,x_2,\cdots, x_n, y_1 ,y_2, \cdots
,y_n)^T$ being a column vector of ${\rr}^{2n}$. Then it is possible
to show~\cite{HOLEVOBOOK} that for any $S \in Sp(2n, \rr)$ there
exists a  {\em canonical} unitary transformation $\hat{U}$ which
maps the canonical observables of the system into a linear
combination of  the operators $\hat{R}_j$, satisfying the condition
\begin{eqnarray}
    \hat{U}^\dag \, \hat{V}(z) \, \hat{U} = \hat{V}(S z)\,, \label{Neq}
\end{eqnarray}
for all $z$. This is often referred to as metaplectic representation.
Conversely, one can show that any unitary $\hat{U}$
which transforms $\hat{V}(z)$ as in Eq.~(\ref{Neq}) must correspond to
an $S\in Sp(2n, \rr)$.

Weyl operators allow one to rewrite the canonical  commutation
relations as
\begin{eqnarray}
\hat{V}(z) \hat{V}(z^\prime)  = \exp[ - \tfrac{i}{2} z^T
\sigma_{2n}   z^\prime ] \hat{V}(z + z^\prime)\,,
\end{eqnarray}
and permit a complete descriptions of the system  in terms of
(characteristic) complex functions. Specifically, any trace-class
operator $\hat{\Theta}$ (in particular, any density operator) can be
expressed as
\begin{eqnarray} \label{NREP1}
\hat{\Theta} = \int \frac{d^{2n} z}{(2\pi)^n}\,
\phi(\hat{\Theta}; z) \, \hat{V}(-z) \,,
\end{eqnarray}
where $d^{2n} z := dx_1 \cdots dx_n dy_1 \cdots dy_n$ and
$\phi(\hat{\Theta}; z)$ is the characteristic function associated
with the operator $\hat{\Theta}$ defined by
\begin{eqnarray}\label{NREP2}
\phi(\hat{\Theta}; z) := \Tr[ \, \hat{\Theta}\, \hat{V}(z) \, ]\,.
\end{eqnarray}
Within this framework a density operator $\hat{\rho}$ of the $n$
modes is said to represent a {\em Gaussian state} if its
characteristic function $\phi(\hat{\rho}; z)$ has a Gaussian form,
i.e.,
\begin{eqnarray}
\phi(\hat{\rho};z) = \exp[ - \tfrac{1}{4} \, z^T \gamma
z + i m^T z] \label{Ngauss1} \,,
\end{eqnarray}
with $m$ being a real vector of mean values $m_j := \Tr [\,
\hat{\rho} \, \hat{R}_j   ]$, and the $2n\times 2n$ real symmetric
matrix $\gamma$ being the {\em covariance
matrix}~\cite{HOLEVOBOOK,EW,FOP} of $\hat{\rho}$. For generic
density operators $\hat{\rho}$  (not only the Gaussian ones) the
latter is defined as the variance of the canonical
 coordinates $\hat{R}$, i.e.,
\begin{eqnarray}
\gamma_{j,j^\prime}  &:=& \Tr\Big[ \rho \big\{ (R_j-m_j) ,
(R_{j^\prime}-m_{j^\prime}) \big\} \Big] \,,
\end{eqnarray}
with $\{ \cdot,\cdot \}$ being the anti-commutator, and it is
bound to satisfy the uncertainty relations
\begin{eqnarray}
\gamma \geqslant i \sigma_{2n}\,, \label{Nuncertainty}
\end{eqnarray}
with $\sigma_{2n}$ being the commutation matrix~(\ref{NSYMP}). Up to an
arbitrary vector $m$, the uncertainty  inequality presented above
uniquely characterizes the set of Gaussian states, i.e. any $\gamma$
satisfying~(\ref{Nuncertainty}) defines a Gaussian state.  Let us
first notice that if $\gamma$ satisfies~(\ref{Nuncertainty})  then
it must be (strictly) positive definite $\gamma >0$, and have $\Det
[\gamma] \geqslant 1$. From Williamson theorem~\cite{Williamson} it
follows that there exists a symplectic $S \in Sp(2n, \rr)$ such that
\begin{eqnarray}\label{Ngammasimp}
\gamma = S      \left[
    \begin{array}{cc}
    D &  {0} \\
    {0} & D
    \end{array}
    \right]
 S^T  \,,
\end{eqnarray}
where  $D := \mbox{diag}( d_1, \cdots, d_n) $ is a diagonal matrix
formed by the {\em  symplectic eigenvalues} $d_j \geqslant 1$  of
$\gamma$. For $S =  \openone_{2n}$ Eq.~(\ref{Ngammasimp}) gives the
covariance matrix associated with thermal bosonic states. This also
shows that any covariance matrix $\gamma$
satisfying~(\ref{Nuncertainty})   can be written as
\begin{eqnarray}\label{Ngammasimp11}
\gamma = S    S^T  + \Delta \,,
\end{eqnarray}
with $\Delta \geqslant 0$\footnote{This is indeed
the matrix
\begin{equation}
\Delta := S
\left[
    \begin{array}{cc}
    D -\openone_n&  {0} \\
    {0} & D -\openone_n
    \end{array}
    \right]
 S^T
\end{equation}
with $D$ as in Eq.~(\ref{Ngammasimp})
 which is positive since $D\geqslant \openone_n$.}.
The extremal solutions of Eq.~(\ref{Ngammasimp11}), i.e.,
$\gamma = S S^T$,  are
{\it minimal uncertainty solutions} and correspond
to the {\it pure Gaussian states} of $n$
modes (e.g., multi-mode squeezed vacuum states).
They are uniquely
determined by the condition $\Det [ \gamma] =1$
and satisfy the condition~\cite{Lindblad}
\begin{eqnarray}
\gamma = - \sigma_{2n} ( \gamma^{-1} ) \sigma_{2n} \label{minimal} \,.
\end{eqnarray}

\subsection{Bosonic Gaussian channels} \label{section12}

In the Schr{\"o}dinger picture evolution is described by applying
the transformation to the states (i.e., the density operators),
$\hat{\rho} \mapsto \Phi(\hat{\rho})$. In the Heisenberg picture the
transformation is applied to the observables of the system, while
leaving the states unchanged, $\hat{\Theta} \mapsto
\Phi_H(\hat{\Theta})$. The two pictures are related through the
identity $\Tr [ \Phi(\hat{\rho}) \hat{\Theta}  ] = \Tr [ \hat{\rho}
\Phi_H(\hat{\Theta})  ],$ which holds for all $\hat{\rho}$ and
$\hat{\Theta}$. The map $\Phi_H$ is called the {\em dual} of $\Phi$.

Due to the representation~(\ref{NREP1}) and (\ref{NREP2}) any
completely positive, trace preserving (CPT) transformation on the
$n$-modes can be characterized by its action on the Weyl operators
of the system in the Heisenberg picture (e.g., see Ref.~\cite{CGH}).
In particular, a {\it bosonic Gaussian channel} (BGC) is defined as
a map which, for all $z$, operates on $V(z)$ according to~\cite{HW}
\begin{equation}\label{d1}
  \hat{V}({z})\longmapsto   \Phi_H( \hat{V}({z}) ) :=
  \hat{V}({Xz})  \, \exp[{- \tfrac{1}{4} z^T Y z } + i v^T z] \,,
\end{equation}
with $v$ being some fixed real vector of $\rr^{2n}$, and
with  $Y, X\in \rr^{2n\times 2n}$ being some fixed real $2n
\times 2n$ matrices satisfying the complete positivity condition
\begin{equation}\label{cptcond}
Y \geqslant i  \Sigma\,   \qquad \mbox{with} \quad \Sigma :=
\sigma_{2n}  - X^{T} \sigma_{2n} X  \,.
\end{equation}
In the context of BGCs the above inequality is the quantum channel
counterpart of the uncertainty relation~(\ref{Nuncertainty}).
Indeed up to a vector $v$,   Eq.~(\ref{cptcond}) uniquely
determines the set of BGCs and bounds $Y$ to be
positive-semidefinite, $Y\geqslant 0$. However, differently
from~(\ref{Nuncertainty}) in this case strict positivity is not a
necessary prerequisite for $Y$. A completely positive map defined
by Eqs.\ (\ref{d1}) and (\ref{cptcond}) will be referred to as
bosonic Gaussian channel (BGC).  As mentioned before, such a map
is a model for a wide class of physical situations, ranging from
communication channels such as optical fibers, to open quantum
systems, and to dynamics in harmonic lattice systems. Whenever one
has only partial access to the dynamics of a system that can be
well-described by a time evolution governed by a Hamiltonian that
is a quadratic polynomial in the canonical coordinates, one will
arrive at a model described by Eqs.\ (\ref{d1}) and
(\ref{cptcond})\footnote{This set does not contain ideal Gaussian
measurements, like optical homodyning~\cite{M}.}.

An important subset of the BGCs is given by set of {\em Gaussian
unitary} transformations which have  $Y=0$, $X \in Sp(2n,
\rr)$, and $v$ arbitrary. They include the canonical
transformations of Eq.~(\ref{Neq}) (characterized by $v=0$), and
the displacement transformations (characterized by having
$X=\openone_{2n}$ and $v$ arbitrary).
The latter simply adds a phase to
the Weyl operators and correspond to unitary transformations of
the form $\Phi_H( \hat{V}({z}) ) :=  \hat{V}(-v) \hat{V}(z)
\hat{V}(v) = \hat{V}({z})  \, \exp[ i v^T z] $.

In the Schr{\"o}dinger picture the BGC transformation~(\ref{d1})
 induces a mapping of the characteristic functions of the form
\begin{equation}
\phi(\hat{\rho};z) \longmapsto \phi(\Phi(\hat{\rho}); z) :=
\phi(\hat{\rho}; Xz)   \, \exp[{- \tfrac{1}{4} z^T Y
z } + i v^T z] \label{NChannel} \,,
\end{equation}
which in turn yields the following transformation of the mean and
the covariance matrix
\begin{eqnarray}
m &\longmapsto& X^T m + v \,, \nonumber \\
\gamma &\longmapsto&    X^T \gamma X +Y
\label{Channel} \,.
\end{eqnarray}
Clearly, BGCs always map Gaussian input states into Gaussian output
states.

For purposes of assessing quantum or classical information
capacities, output entropies, or studying degradability or
anti-degradability of a channel \cite{CG,CGH,HOLEVONEW,WGG}, the
full knowledge of the channel is not required: Transforming the
input or the output  with any  unitary operation (say, Gaussian
unitaries) will not alter any of these quantities. It is then
convenient to take advantage of this freedom to simplify the
description of the BGCs. To do so we first notice that the set of
Gaussian maps is closed under composition.  Consider then
$\Phi^\prime$ and $\Phi^{\prime\prime}$ two BGCs described
respectively by the elements $X^\prime, Y^\prime, v^\prime$ and
$X^{\prime\prime}, Y^{\prime\prime}, v^{\prime\prime}$. The
composition $\Phi^{\prime\prime} \circ \Phi^\prime$ where, in
Schr\"{o}dinger representation, we first operate with $\Phi^\prime$
and then with $\Phi^{\prime\prime}$,  is still a BGC and it is
characterized by the parameters
\begin{eqnarray}
v &=& (X^{\prime\prime} )^T v^\prime + v^{\prime\prime}\,, \nonumber \\
X &=& X^\prime  X^{\prime\prime}\nonumber \,, \\
Y &=& (X^{\prime\prime})^T\,  Y^\prime\,  X^{\prime\prime} +
Y^{\prime\prime}\,.\label{NCOMP}
\end{eqnarray}
Exploiting these composition rules it is then easy to verify that
the vector $v$ can always be compensated by properly displacing
either the input state or the output state (or both)   of the
channel. For instance by taking $X^{\prime\prime} =
\openone_{2n}$, $Y^{\prime\prime} =0$ and $v^{\prime\prime} = -
v^\prime$, Eq.~(\ref{NCOMP}) shows that $\Phi^\prime$ is unitarily
equivalent to the Gaussian channel $\Phi$ which has $v =0$ and $X
= X^\prime$, $Y= Y^\prime$. Therefore, without loss of generality,
in the following we will focus on BGCs having $v=0$.

More generally consider the case where we
cascade a generic BGC $\Phi^\prime$ described by matrices $X^\prime,
Y^\prime$ as in Eq.~(\ref{cptcond}) with a couple of canonical
unitary transformation $\hat{U}_1$ and $\hat{U}_2$ described by the
symplectic matrices $S_1$ and $S_2$ respectively. The resulting BGC
$\Phi$ is then described by the matrices
\begin{eqnarray}\label{NXY}
X &=& S_1
(X^\prime) S_2, \, \\
Y &=& S_2^T  (Y^\prime) S_2 \nonumber \,.
 \end{eqnarray}
For single mode ($n = 1$) this procedure  induces a simplified
canonical form~\cite{SEW,CGH,HOLEVONEW} which, up to a Gaussian
unitarily equivalence, allows one to focus only on transformations
characterized by $X$ and $Y$  which, apart from some special cases,
are proportional to the identity. In this paper we will generalize
some of these results to an arbitrary number of modes $n$. To
achieve this goal, in the following section we  first present an
explicit dilation representation in which the mapping~(\ref{d1}) is
described as  a (canonical) unitary coupling between the $n$ modes
of the system and some extra {\em environmental} modes which are
initially prepared into a Gaussian state. Then we will introduce the
notion of minimal noise channel, showing a useful decomposition
rule.

 \section{Unitary dilation theorem} \label{sec:unitary}

In this section we introduce a general construction of unitary
dilations of multi-mode quantum channels. Specifically we show
that a CPT channel acting on $n$ modes is a BGC if and
only if it can be realized by invoking $\ell \leqslant 2n$
additional (environmental) modes $E$ through the expression
\begin{equation}
    \Phi(\hat{\rho})
=    \text{Tr}_E[\hat{U} (\hat{\rho} \otimes \hat{\rho}_E)\hat{U}^\dagger]\,,
    \label{eq1}
\end{equation}
where $\hat{\rho}$ is the input $n$-mode state of the system,
$\hat{\rho}_E$ is a Gaussian state of an environment, $\hat{U}$ is a
canonical unitary transformation which couples the system with the
environment, and $\text{Tr}_E$ denotes the partial trace over $E$.
In case in which $\hat{\rho}_E$ is  pure,
Eq.~(\ref{eq1}) corresponds to  a Stinespring  dilation~\cite{STINE}
of the channel $\Phi$, otherwise it is a physical representation
analogous to those employed in Refs.~\cite{CG,CGH} for the single
mode case.

\subsection{General dilations}

In this subsection, we will construct Gaussian dilations, including
a discussion of all rank-deficient cases, and will later focus on
dilations involving the minimal number of modes. To proceed, we will
first establish some conventions and notation. To start with, we
write the commutation matrix of our $n + \ell$ modes in the block
structure
 \begin{equation}
    {\sigma} :=  \sigma_{2n} \oplus \sigma^{E}_{2\ell} =
     \left[
    \begin{array}{cc}
     \sigma_{2n}& {0} \\
    {0} & \sigma^{E}_{2\ell}    \end{array}
    \right]
    \begin{array}{l}
    \} \, 2 n \\
    \} \, 2 \ell \,,
      \end{array}
    \label{NSYMPnew}
\end{equation}
where  $\sigma_{2n}$ and $\sigma^{E}_{2\ell}$  are $2n\times2n$
and $2\ell \times 2\ell$ commutation matrices associated with the
system and environmental modes, respectively. For $\sigma_{2n}$ we
assume the structure as defined in Eq.\
(\ref{NSYMP}). For $\sigma^{E}_{2\ell}$,
in contrast, we do not make any assumption at this point,
leaving open the possibility
of defining it later on\footnote{With this choice the canonical
commutation relations of the $n+\ell$ mode read as $[\hat{R}_j,
\hat{R}_{j^\prime}] = i {\sigma}_{j,j^\prime}$ where
$\hat{R} := ( \hat{Q}_1, \cdots, \hat{Q}_n; \hat{P}_1, \cdots,
\hat{P}_n;  \hat{r}_1,  \cdots, \hat{r}_{2\ell} )$ with
$\hat{Q}_j, \hat{P}_j$  being the canonical coordinates of the
$j$-th system mode and  with and $\hat{r}_1, \cdots,
\hat{r}_{2\ell}$ being some ordering of the canonical
coordinates $\hat{Q}_1^E, \hat{P}_1^E; \cdots
 ; \hat{Q}_\ell^E, \hat{P}_\ell^E$ of the
 environmental modes. For instance, taking
$\sigma_{2\ell}^E = \sigma_{2\ell}$ corresponds to
have $\hat{R}
:= ( \hat{Q}_1, \cdots, \hat{Q}_n; \hat{P}_1, \cdots,
\hat{P}_n;  \hat{Q}_1^E,  \cdots, \hat{Q}_\ell^E ;
\hat{P}_1^E,\cdots , \hat{P}_{\ell}^E )$.}. Accordingly, the
canonical unitary transformation $\hat{U}$ of Eq.~(\ref{eq1}) will
be uniquely determined by a $2(n + \ell) \times 2(n + \ell)$ real
matrix $S\in Sp(2(n+\ell),\rr)$ of block form
 \begin{equation}
   S :=
     \left[
    \begin{array}{cc}
   s_1& s_2 \\
    s_3 & s_4    \end{array}
    \right] \label{NmatrixS} \,,
\end{equation}
which satisfies the condition
\begin{eqnarray}  \label{NSN}
S  {\sigma} S^T = {\sigma} \,, \qquad \Longleftrightarrow  \quad
\left\{ \begin{array}{l}
s_1 \, \sigma_{2n} \, s_1^T + s_2 \, \sigma^{E}_{2\ell} \, s_2^T =\sigma_{2n} \,, \\
\\
s_1 \, \sigma_{2n} \,  s_3^T + s_2 \, \sigma^{E}_{2\ell} \, s_4^T =0 \,,
\\ \\
s_3 \, \sigma_{2n} \, s_3^T + s_4 \, \sigma^{E}_{2\ell} \,s_4^T =\sigma^{E}_{2\ell} \,.
\end{array}
\right.
\end{eqnarray}
In the above expressions, $s_1$ and  $s_4$ are  $2n\times2n$ and
$2\ell \times 2\ell$ real square matrices, while $s_2$ and
$s_3^T$ are $2 n \times 2\ell$ real  rectangular matrices.
Introducing then the covariance matrices $\gamma\geqslant i
\sigma_{2n}$ and $\gamma_E \geqslant i\sigma^{E}_{2\ell}$ of the
states $\hat{\rho}$ and $\hat{\rho}_E$, the identity~(\ref{eq1})
can be written as
\begin{equation}\label{dil}
    \left. S   \left[
    \begin{array}{cc}
   \gamma & 0 \\
    0 & \gamma_E   \end{array}
\right] S^{T} \right|_{2n}= s_1 \, \gamma \, s_1^T + s_2 \,
\gamma_E \, s_2^T = X^{T}\gamma X +Y,
\end{equation}
where $|_{2n}$ denotes the upper principle submatrix of degree
$2n$, and where  $X, Y\in  \rr^{2n\times 2n}$ satisfying the
condition~(\ref{cptcond}) are the matrices associated with the
channel $\Phi$. In writing Eq.~(\ref{dil}) we use the fact that
due to the definition~(\ref{NSYMPnew}) the covariance matrix of
the composite state $\hat{\rho} \otimes\hat{\rho}_E$ can be
expressed as $\gamma\oplus \gamma_E$. With these definitions, the
first part of the unitary dilation property~(\ref{eq1}) can  be
written as follows:

\begin{prop}[Unitary dilations of Gaussian channels]
Let $\gamma_E$ be the covariance matrix of a Gaussian state of
$\ell$ modes and let $S\in Sp(2(n+\ell),\rr)$ be a symplectic
transformation. Then there exists a symmetric $2n\times 2n$-matrix
$Y\geqslant 0$ and a $2n\times 2n$-matrix $X$ satisfying the
condition (\ref{cptcond}), such that Eq.~(\ref{dil}) holds for all
$\gamma$.
\end{prop}

{\it Proof:} The proof is straightforward: We write $S$ in the block
form~(\ref{NmatrixS}) and take $X= s_1^T$ and $Y= s_{2}\gamma_E
s_{2}^{T}$. Since $\gamma_E$ is a covariance matrix of $\ell$ modes,
$\gamma_E - i\sigma_{2\ell}\geqslant 0$  and therefore
$s_{2}(\gamma_E - i\sigma_{\ell})s_{2}^{T}\geqslant 0$. This leads
to Eq.~(\ref{cptcond}) through the identity the symplectic condition
 $s_1 \sigma_{2n} s_1^T + s_2 \sigma_{2\ell} s_2^T =
\sigma_{2n}$ which follows by comparing the upper principle
submatrices of degree $n$  of both terms of Eq.~(\ref{NSN}).
$\blacksquare$
\newline

This proves that any CPT map obtained by coupling the $n$ modes with
a Gaussian state of  $\ell$ environmental bosonic modes through a
 Gaussian unitary $\hat U$ is a BGC. The converse property is
more demanding. In order to present it we find it useful to state
first the following

\begin{lemma}[Extensions of symplectic forms]
Let, for some skew symmetric $\sigma^{E}_{2\ell}$, $s_1$ and $s_2$
be $2n\times 2n$ and $2n\times 2\ell$ real matrices forming a
symplectic system, i.e., $s_1 \, \sigma_{2n} \, s_1^T + s_2 \,
\sigma^{E}_{2\ell} \, s_2^T =\sigma_{2n}$. Then we can always find
real matrices $s_3$ and $s_4$ such that $S$ of Eq.~(\ref{NmatrixS})
is symplectic with respect to the commutation matrix~(\ref{NSYMPnew}).
\end{lemma}
{\it Proof:} Since the rows of $S$ form a symplectic basis, given
$s_1$ and $s_2$ (an incomplete symplectic basis), it is always
possible to find $s_3$ and $s_4$ as above. The proof easily follows
from a skew-symmetric version of the Gram-Schmidt process to
construct a symplectic basis~\cite{silva}. For a special subset of
BGCs,  in Sec.~\ref{sec:canonical} we will present an explicit
expression for $S$ based on a simplified (canonical) representation
of the $X$ matrix that defines $\Phi$. See also
Appendix~\ref{appendixA}.~$\blacksquare$
\newline

Due to the above result, the possibility of realizing unitary
dilation Eq.~(\ref{eq1}) for a generic BGC described by the matrices
$X$ and $Y \geqslant i \Sigma = i (\sigma_{2n} - X^T \sigma_{2n}
X)$, can be proven by simply taking  $s_1 = X^T$ and finding some
$2n\times 2\ell$ real matrix $s_2$ and an $\ell$-mode covariance
matrix $\gamma_E \geqslant i \sigma_{2\ell}^E$ that solve  the
equations
 \begin{eqnarray}
    s_2 \, \sigma^{E}_{2\ell} \, s_2^T &=& \sigma_{2n} -
    s_1 \, \sigma_{2n} \,
    s_1^T = \Sigma \,, \label{probb2} \\
    s_2 \, \gamma_E \, s_2^T &=& Y \label{probb1} \,.
\end{eqnarray}
With this choice in fact Eq.\ (\ref{dil}) is trivially satisfied
for all $\gamma$, while $s_1$ and $s_2$ can be completed to a
symplectic matrix $S\in Sp(2(n+ \ell), \rr)$. The unitary dilation
property~(\ref{eq1}) can  hence be restated as follows:

\begin{theorem}[Unitary dilations of Gaussian channels: Converse
implication]

For
\newline
any real $2n\times 2n$-matrices $X$ and $Y$ satisfying the
condition~(\ref{cptcond}), there exist $\ell$ smaller than or
equal to $2n$, $S\in Sp(2(n+\ell),\rr)$, and a covariance
matrix $\gamma_E$ of $\ell$ modes, such that Eq.~(\ref{dil}) is
satisfied.
\end{theorem}

{\it Proof:} As already noticed the whole problem can be solved by
assuming $s_1 = X^T$ and finding $s_2$ and $\gamma_E$ that satisfy
Eqs.~(\ref{probb2}) and (\ref{probb1}).
We start by observing that the $2n\times 2n$ matrix $\Sigma$ defined
in Eq.~(\ref{cptcond}) is skew-symmetric, i.e., $\Sigma = -
\Sigma^T$. Moreover according to Eq.~(\ref{cptcond}) its support
must be contained in the support of $Y$, i.e.,
$\mbox{Supp}[\Sigma] \subseteq \mbox{Supp}[Y]$. Consequently given
$k := \mbox{rank}[Y]$ and  $r := \mbox{rank}[\Sigma]$
as the ranks of $Y$ and $\Sigma$, respectively, one
has that $k \geqslant r$.
We can hence identify three different regimes:
\begin{itemize}
\item[(i)] $k=2n$, $r=2n$,
i.e.,  both $Y$ and $\Sigma$ are full rank and hence invertible.
Loosely speaking, this means that all the noise components in the
channel are basically quantum (although may involve classical noise
as well).
\item[(ii)] $k=2n$ and $r <2n$,
i.e., $Y$ is full rank and hence invertible, while $\Sigma$ is
singular. This means that the some of the noise components can be
purely classical, but still nondegenerate.
\item[(iii)] $2n > k \geqslant r$,
i.e., both $Y$ and $\Sigma$ are singular. There are noise components
with zero variance.
\end{itemize}
Even though (i)  and (ii) admit similar solutions, it is
instructive to analyze them separately. In the former case in fact
there is a  simple direct way of  constructing a physical
dilation of  the channel with $\ell
=n$  environmental modes.

{(i)}  Since $\Sigma$ is skew-symmetric and invertible  there
exists an $O\in O(2n, \rr)$ orthogonal such that
   \begin{eqnarray}
    O \Sigma O^{T}&=& \left[
    \begin{array}{cc}
    \,\,\,{0} \,\,\,&  \mu  \\
    - \mu  & \, \,\,{0}\,\,\,
    \end{array}
    \right] \,, \label{Nimpo1}
    \end{eqnarray}
    where $\mu =\text{diag}(\mu_1,\cdots, \mu_n)$ and $\mu_{i}>0 $ for
all $i=1,\cdots,n$ (see page 107 in Ref.\
\cite{HJ}). Hence $K:=
M^{-1/2} O$ with $M:= \mu \oplus \mu$ satisfies
 \bea \label{neweq1}
 K \Sigma K^T = \sigma_{2n} \,.
 \eea
Taking then $s_2 := K^{-1}$ we get\footnote{From now on, the
symbol $A^{-T}$ will be used to indicate the transpose of the
inverse of the matrix $A$, i.e.,
 $A^{-T} := (A^{-1})^T = (A^{T})^{-1}$.}
\begin{eqnarray}
s_2\, \sigma_{2n} \,s_2^T =
 K^{-1}\,  \sigma_{2n} \, K^{-T} =
\Sigma \,, \label{afafa}
\end{eqnarray}
which corresponds to Eq.~(\ref{probb2}) for $\ell =n$. Since $s_1
= X^T$, Lemma 1 guarantees that this is sufficient to prove the
existence of $S$. The condition~(\ref{dil}) finally follows by
taking $\gamma_E= K Y K^T$ which is strictly positive (indeed $K$
is invertible and $Y>0$ because it has full rank) and
which satisfies the
uncertainty relation~(\ref{Nuncertainty}), i.e.,
\begin{eqnarray}\label{partisolution}
Y \geqslant i \Sigma \quad  \Longrightarrow \quad \gamma_E =K Y K^T
\geqslant i K \Sigma K^T = i \sigma_{2n} \,. \end{eqnarray}

This shows that the channel admits a unitary dilation of the
form as specified in Eq.\ (\ref{eq1})
with  $\ell = n$ environmental modes with commutation matrix,
$\sigma^E_{2n} = \sigma_{2n}$ -- see discussion after
Eq.~(\ref{NSYMPnew}). Such a solution, however,
will involve a  pure
state $\hat{\rho}_E$  only if $\Det[ \gamma_E] =1$, i.e.,
\begin{eqnarray}\label{puritycond}
\Det[ Y ] \Det[ K]^2 =1 \quad  \Longleftrightarrow \quad   \Det[Y] = \Det [\Sigma] \,.
\end{eqnarray}
When $\Det[ \gamma_E] > 1$, i.e., $\Det[Y]  >  \Det[ \Sigma]$, we
can still construct a pure dilation by simply adding further $n$
modes which purify the state associated with the covariance
matrix $\gamma_E$ and by extending the unitary operator $\hat{U}$
associated with $S$ as the identity operator on them.
For details see the discussion of case (ii) given below. This
corresponds to constructing a unitary dilation~(\ref{eq1}) with
the pure
state $\hat{\rho}_E$ being defined on $\ell =2n$ modes.
\newline

(ii) In this case $Y$ is still invertible, while $\Sigma$ is not.
Differently from the approach we adopted in solving case (i),
we here derive directly a Stinespring unitary dilation, i.e.,  we
construct a solution with a pure $\gamma_E$
that involves $\ell =2n$ environmental modes.
In the next section, however, we will show
that, dropping the purity requirement, one can construct unitary
dilation that involves $\hat{\rho}_E$ with only $\ell = 2n -r/2$
modes.

To find $s_2$ and $\gamma_E$ which solve Eqs.~(\ref{probb2}) and
(\ref{probb1}), it is useful to first transform $Y$ into a simpler
form by a congruent transformation, i.e.,
 \begin{eqnarray}
C Y C^T = \openone_{2n} \,, \label{Ncongruent}
\end{eqnarray}
with  $C\in Gl(2n, \rr)$ being not singular, e.g.,
$C := Y^{-1/2}$.
From Eq.~(\ref{cptcond}) it then follows that
\begin{equation}\label{cptcond1}
    \openone_{2n}  \geqslant i  \Sigma^\prime \,,
    \end{equation}
with $\Sigma^\prime :=  Y^{-1/2}\,  \Sigma \, Y^{-1/2}$ being
skew-symmetric (i.e., $\Sigma^\prime = -(\Sigma^\prime)^T$)
and
singular with
$\mbox{rank}[ \Sigma^\prime ] = \mbox{rank}[ \Sigma]=r$~\cite{HJ}.
  We then observe that introducing
  \begin{eqnarray}
  s_2 = Y^{1/2}\,  s_2^\prime \,, \label{NdefS2prime}
  \end{eqnarray}
the conditions~(\ref{probb2}) and (\ref{probb1}) can  be written
as
\begin{eqnarray}
    s_2^\prime \, \sigma_{2\ell}^E
    \,(s_2^\prime)^T &=& \Sigma^\prime\,, \label{prob2} \\
    s_2^\prime \,  \gamma_E  \, (s_2^\prime)^T &=&
    \openone_{2n}\,.\label{prob1}
\end{eqnarray}
Finding $s_2^\prime$ and $\gamma_E$ which satisfy these
expressions will provide us also a solution of Eqs.~(\ref{probb2})
and (\ref{probb1}).

As in the case of Eq.~(\ref{Nimpo1}), there exists an orthogonal
matrix $O \in O(2n, \rr)$ which transforms the
skew-symmetric matrix $\Sigma^\prime$ in a  simplified block form.
In this case however, since $\Sigma^\prime$ is singular, we
find~\cite{HJ}
\begin{eqnarray}  \label{Mequ1}
O \Sigma^\prime O^{T}=
     \left[
    \begin{array}{c|c}
    {0} &  \begin{array}{c|c}
    \mu  &  0 \\ \hline
    \;\;0 \;\;& \;\; {0}\;\;
    \end{array}
     \\ \hline
     \begin{array}{c|c}
    - \mu& 0 \\ \hline
    \;\;0 \;\; &\;\;{0}\;\;
    \end{array}
     & {0}
    \end{array}
    \right] \begin{array}{l}
    \} \, r/2 \\
    \} \, n-r/2\\
     \} \, r/2 \\
    \} \, n-r/2 , \\
    \end{array}
    \end{eqnarray}
where now $\mu= \mbox{diag}(\mu_1,\cdots, \mu_{r/2})$ is the
$r/2\times r/2$  diagonal matrix formed by the strictly
positive
eigenvalues of $|\Sigma^\prime|$ which satisfy the conditions
$ 1\geqslant \mu_j>0$, this being  equivalent with
\begin{equation}
    \openone_{r/2} \geqslant \mu,
     \label{Ncondizioneimpo}
\end{equation}
as a consequence of inequality~(\ref{cptcond1}).
Define then  $K:= M^{-{1}/{2}} \, O$ with
 \begin{eqnarray} M =
\left[
    \begin{array}{c|c}
    \begin{array}{c|c}
    \mu  & \;\;\;\; 0 \;\; \;\;\\ \hline
    \;\;\;\;0\;\;\;\; & \openone_{n-r/2}
    \end{array} &  0
     \\ \hline
    0
     &  \begin{array}{c|c}
    \mu & \;\;\;\; 0 \;\; \;\; \\ \hline
    \;\;\;\; 0 \;\; \;\; & \openone_{n-r/2}
    \end{array}    \end{array}
    \right] \begin{array}{l}
    \} \, r/2 \\
    \} \, n-r/2\\
     \} \, r/2 \\
    \} \, n-r/2 . \\
    \end{array}
    \end{eqnarray}
It  satisfies the identity
\begin{eqnarray} K \Sigma^\prime K^{T}&=& \left[
    \begin{array}{c|c}
    {0} & \begin{array}{c|c}
    \openone_{r/2} &  \;\;0\;\; \\   \hline   \;\;0\;\; &  {0}
    \end{array}
     \\ \hline
     \begin{array}{c|c}
    - \openone_{r/2} & \;\;0\;\; \\ \hline
    \;\;0\;\; & {0}
    \end{array}
    &  {0}
    \end{array}
    \right] \begin{array}{l}
    \} \, r/2 \\
    \} \, n-r/2\\
     \} \, r/2 \\
    \} \, n-r/2 . \\
    \end{array} \label{NIDENTITY}
    \end{eqnarray}
To show that Eqs.~(\ref{prob2}) and (\ref{prob1}) admit a
solution we
take $\ell =2 n$ and write $\sigma_{4n}^E = \sigma_{2n} \oplus
\sigma_{2n}=\sigma_{4n}$
with  $\sigma_{2n}$ as in Eq.~(\ref{NSYMP}). With
these definitions the $2n\times 4n$ rectangular matrix
$s_2^\prime$ can be chosen to have the block structure
\begin{eqnarray} \label{defs2prime}
s^\prime_2 &=& \left[
    \begin{array}{c|c}
    K^{-1} & O^T A
    \end{array}
    \right] \,,
    \end{eqnarray}
    with $A$ being the following $2n\times 2n$ symmetric matrix
    \begin{eqnarray}\label{defA} A =  A^T &=& \left[
    \begin{array}{c|c}
    {0} & \begin{array}{c|c}
    \;\;\;\;0\;\;\;\; & \;\;\;\;0\;\;\;\; \\  \hline
    0 & \openone_{n-{r}/{2}}
    \end{array}
        \\ \hline
     \begin{array}{c|c}
    \;\;\;\;0\;\;\;\; & \;\;\;\;0\;\;\;\; \\ \hline
    0 & \openone_{n-r/2}
    \end{array}
    & {0}
    \end{array}
    \right] \begin{array}{l}
    \} \, r/2 \\
    \} \, n-r/2\\
     \} \, r/2 \\
    \} \, n-r/2 .\\
    \end{array}\,
    \end{eqnarray}
By direct substitution  one can easily verify that
Eq.~(\ref{prob2}) is indeed satisfied, see
Appendix~\ref{AppendixB} for details. Inserting
Eq.~(\ref{defs2prime})
 into Eq.~(\ref{prob1})
 yields now the following equation
 \begin{eqnarray} \label{NNequation}
 \alpha + A\,  \delta^T + \delta \, A^T + A \, \beta \, A^T = M^{-1} \,,
 \end{eqnarray}
for the $4n\times 4n$ covariance matrix
\begin{eqnarray}
\gamma_E = \left[
    \begin{array}{cc}
    \alpha & \, \delta \\
    \delta^T & \beta
     \end{array}
    \right]
    \label{gammaedef}\,,
    \end{eqnarray}
see Appendix~\ref{AppendixC} for details.
       A solution can be easily derived by taking
    \begin{eqnarray}
    \alpha =\beta =
     \left[
    \begin{array}{c|c}
        \begin{array}{c|c}
    \mu^{-1} & \;\;\;\;0\;\;\;\; \\ \hline
    \;\;\;\; 0\;\;\;\;  &  \xi \openone_{n-r/2}
    \end{array}
& 0 \\ \hline
    0 &          \begin{array}{c|c}
    \mu^{-1} & \;\;\;\; 0 \;\;\;\; \\ \hline
    \;\;\;\; 0 \;\;\;\; &  \xi \openone_{n-r/2} \\
    \end{array}
    \end{array}
    \right]
    \begin{array}{l}
    \} \, r/2 \\
    \} \, n-r/2\\
     \} \, r/2 \\
    \} \, n-r/2 , \\
    \end{array}\label{Nalpha}
    \end{eqnarray}
   with $\xi =5/4$ and
    \begin{eqnarray}
    \delta = \left[
    \begin{array}{c|c}
        0
& \begin{array}{c|c}
 f(\mu^{-1}) & \;\;\;\; 0 \;\;\;\; \\ \hline
     \;\;\;\;\;\; 0 \;\;\;\;\;\; &  f(\xi \openone_{n-r/2})
    \end{array} \\ \hline
     \begin{array}{c|c}
    f(\mu^{-1}) & \;\;\;\; 0 \;\;\;\; \\ \hline
   \;\;\;\; \; \;0 \;\;\;\;\;\; &   f(\xi \openone_{n-r/2})
    \end{array}  &  0
    \end{array}
    \right]  \begin{array}{l}
    \} \, r/2 \\
    \} \, n-r/2\\
     \} \, r/2 \\
    \} \, n-r/2 , \\
    \end{array} \label{Ndelta}
    \end{eqnarray}
with $f(\theta) := - ({\theta^2 -\openone})^{1/2}$. For all diagonal
matrices $\mu$ compatible with the
constraint~(\ref{Ncondizioneimpo}) the resulting $\gamma_E$
satisfies the uncertainty relation $\gamma_E \geqslant i
\sigma_{4n}$. Moreover since it has $\Det[ \gamma_E] =1$, this is
also a minimal uncertainty state, i.e., a pure Gaussian state of
$2n$ modes. It is worth stressing that for $r=2n$, i.e., when also
the rank of  $\Sigma$ is maximum, the above solution provides an
alternative derivation of the unitary dilation discussed in the part
(i) of the theorem. In this case the covariance matrix $\gamma_E$
has block elements
     \begin{eqnarray}
    \alpha =\beta =
     \left[
    \begin{array}{cc}
         \mu^{-1}
& 0 \\
    0 &              \mu^{-1}
    \end{array}
    \right]
    \begin{array}{l}
        \} \, n\\
     \} \, n
        \end{array}  \,, \qquad  \quad
               \delta = \left[
    \begin{array}{cc}
        0
&
 f(\mu^{-1})    \\
    f(\mu^{-1})
      &  0
    \end{array}
    \right]  \begin{array}{l}
       \} \, n\\
     \} \, n
    \end{array} \,,  \label{Ndelta1}
    \end{eqnarray}
where $\mu$ is now a strictly positive $n\times n$ matrix, while
Eqs.~(\ref{NdefS2prime}) and (\ref{defs2prime}) yield
   \begin{eqnarray}
   s_2 :=
   Y^{1/2} O^T \left[
    \begin{array}{c|c}
     \begin{array}{c|c}
     \mu^{1/2}    & 0 \\ \hline
     0 & \mu^{1/2}
    \end{array}
     &  \begin{array}{c|c}
    0   & 0  \\ \hline
     0 & 0
    \end{array}
    \end{array}
    \right] \begin{array}{l}
       \} \, n\\
     \} \, n .
    \end{array}
    \end{eqnarray}
       \newline

(iii) Here both $Y$ and $\Sigma$ are singular.
This case is very similar to case (ii). Here, the
dilation can be constructed by introducing a strictly positive
matrix $\bar{Y}>0$ which satisfies the condition
  \begin{eqnarray} \label{defbary}
  \Pi \, \bar{Y} \, \Pi = Y \,,
  \end{eqnarray}
with $\Pi$ being the projector onto the support of $Y$. Such a
$\bar{Y}$ always exists ($\bar{Y} = Y + (\openone- \Pi)$).
By construction, it satisfies the inequality $\bar{Y} \geqslant
Y\geqslant i \Sigma$.  According to Sec.~\ref{section12},
$\bar{Y}$ and $X$ define thus  a BGC. Moreover, since $\bar{Y}$ is
strictly positive, it has full rank. Therefore,  we can  use part
(ii) of the proof to  find a $2n \times 2\ell$ matrix
$\bar{s}_2$ and $\bar{\gamma}_E \geqslant i \sigma_{2\ell}$ which
satisfy the conditions~(\ref{probb2}) and (\ref{probb1}), i.e.
\begin{eqnarray}
\bar{s}_2 \, \sigma^{E}_{2\ell} \, \bar{s}_2^T &=& \Sigma \,, \label{probb22} \\
\bar{s}_2 \, \bar{\gamma}_E \, \bar{s}_2^T &=& \bar{Y}
\label{probb11} \,. \end{eqnarray} A unitary dilation for the
channel $Y, X$ is then obtained by choosing $\gamma_E =
\bar{\gamma}_E$ and $s_2 = \Pi \bar{s}_2$. In fact from
Eq.~(\ref{probb11}) we get
    \begin{eqnarray}
     s_2 \,  \gamma_E\, s_2^T =
    \Pi\,  \bar{s}_2 \, \bar{\gamma}_E \, \bar{s}_2^T\,  \Pi
    = \Pi \, \bar{Y} \, \Pi = Y \,,
    \end{eqnarray}
    while from Eq.~(\ref{probb22})
       \begin{eqnarray}
     s_2 \, \sigma_{2\ell}^E \, s_2^T =
    \Pi\,  \bar{s}_2 \, \sigma_{2\ell}^E\,  \bar{s}_2^T\,  \Pi
    = \Pi\,  \Sigma \,  \Pi   = \Sigma \,,
    \end{eqnarray}
where we have used the fact that $\mbox{Supp}[\Sigma] \subseteq
\mbox{Supp}[Y]$.
   $\blacksquare$
\newline

In proving the second part of the unitary dilations theorem we
provided explicit expressions for the environmental state
$\hat{\rho}_E$ of Eq.~(\ref{eq1}). Specifically such a state is
given by the pure $2n$ mode Gaussian state $\hat{\rho}_E$
characterized by the covariance matrix $\gamma_E$ of
elements~(\ref{Nalpha}) and (\ref{Ndelta}). A trivial observation
is that this can always be replaced by the $2n$ modes vacuum state
$|\O\rangle\langle \O|$ having the covariance matrix
$\gamma_E^{(0)} = \openone_{2n}$. This is a consequence of the
obvious property that according to Eq.~(\ref{Ngammasimp11})
all pure Gaussian
states are equivalent to $|\O\rangle\langle \O|$ up to a
Gaussian unitary transformation. On the level of covariance
matrices, Gaussian unitaries correspond to symplectic
transformations. For a remark on unitarily
equivalent dilations, see also Appendix D. Hence,
by means of a congruence with an
appropriate symplectic transformation, we immediately
arrive at the following corollary:

\begin{corollary}[Gaussian channels with pure Gaussian
dilations]
Any $n$-mode Gaussian channel $\Phi$ admits a Gaussian unitary
dilation~(\ref{eq1}) with $\hat{\rho}_E=|\O\rangle\langle \O|$
being the vacuum state on $2n$ modes.
 \end{corollary}

\subsection{Reducing the number of environmental modes}\label{sec:red}

An interesting question is the characterization of the minimal
number of environmental modes $\ell$  that need to be involved in
the unitary dilation~(\ref{eq1}). From Theorem 1 we know that such
number is certainly smaller than or equal to twice the number $n$ of
modes on which the BGC is operating:  We have in fact explicitly
constructed one of such representations that involves $\ell =2n$
modes in a minimal uncertainty, i.e., pure Gaussian state. We also
know, however, that there are situations\footnote{Not mentioning the
trivial case of Gaussian unitary transformation which does not
require any environmental mode to construct a unitary dilation.} in
which $\ell$ can be reduced to just $n$: This happens for instance
for BGCs $\Phi$ with  $\mbox{rank}[Y] = \mbox{rank}[\Sigma]=2n$,
i.e., case (i)  of Theorem 1. In this case one can  represent the
channel $\Phi$ in terms of a Gaussian unitary coupling with $\ell =
n$ environmental modes which are prepared into a Gaussian state with
covariance matrix
\begin{eqnarray}
\gamma_E= K Y K^T \label{Nimp} \,,
\end{eqnarray}
-- see  Eq.~(\ref{partisolution}).  In general, this will not be
of Stinespring form (not be a pure unitary dilation) since
$\gamma_E$ is not a minimal uncertainty covariance matrix. In
fact, for $n=1$ this corresponds  to the physical representation
of $\Phi$ of Refs.~\cite{CGH}. However if $Y$ and $X$ satisfy the
condition (\ref{puritycond}), our analysis provides a unitary
dilation involving merely $\ell = n$  modes in a pure Gaussian
state.

We can then formulate a necessary and sufficient condition for
the channels $\Phi$ of class (i) which can be described in terms
of $n$ environmental modes prepared into a pure state.
It is given by
\begin{equation}\label{minimalchannel}
    Y=  \Sigma \,Y^{-1}\,  \Sigma^T \,,
\end{equation}
which follows by imposing the minimal uncertainty
condition~(\ref{minimal}) to the $n$-mode covariance
matrix~(\ref{Nimp}) and  by using (\ref{neweq1}). Similarly one
can verify that given  a pure $n$-modes Gaussian
state $\hat{\rho}_E$ and an
$S \in Sp(4n, \rr)$
(\ref{NmatrixS}) with an invertible subblock $s_2$,
then  the  corresponding
BGC satisfies condition~(\ref{minimalchannel}).
The above result can be strengthened by looking at the solutions
for channels of class (ii) of which the channel of class (i) are a
proper subset.

To achieve this goal,
let us first note that with the choice we made on
 $\sigma_{2\ell}^E=\sigma_{4n}$,
the two matrices $\alpha$ and $\beta$ of
Eq.~(\ref{Nalpha}) are $2n\times 2n$ covariance matrices for two
sets of independent $n$ bosonic modes satisfying the
uncertainty relations~(\ref{Nuncertainty})  with respect to the form
$\sigma_{2n}$. In turn, the matrices
$\delta$ and $\delta^T$ of Eq.~(\ref{Ndelta})
represent cross-correlation terms among such sets. After all,
the entire covariance matrix $\gamma_E$ corresponds
to a pure Gaussian state.

They key point is now the observation that in
Eq.~(\ref{NNequation}), the matrix $A$ couples only with those
rows and columns of the matrices $\delta$ and $\beta$ which
contain elements $\xi \openone_{n-r/2}$ or
$f(\xi\openone_{n-r/2})$: As far as $A$ is concerned, one could
indeed replace the element $\mu^{-1}$ and $f(\mu^{-1})$ of such
matrices with zeros. The only reason we keep these elements, in
the way they are in Eqs.~(\ref{Nalpha}) and (\ref{Ndelta}), is to
render $\gamma_E$ the covariance matrix of a minimal uncertainty
state. In other words,  the elements of $\delta$ and $\beta$
proportional to $\mu^{-1}$ or $f(\mu^{-1})$ are only introduced to
purify the corresponding elements of the submatrix $\alpha$, which
is in itself hence a covariance matrix of a mixed Gaussian state.

Suppose then that   $\mu$ of Eq.~(\ref{Ncondizioneimpo}) has (say)
the first $r^\prime/2$ eigenvalues equal to $1$, i.e., $\mu_{1} =
\mu_{2} = \cdots = \mu_{r^\prime/2} =1$ while for $j \in \{
r^\prime/2 +1, \cdots, r/2\}$ we have that
 $\mu_{j} \in (0,1)$. In this case
the corresponding sub-matrix of $\alpha$ associated with  those
elements represent a pure Gaussian state, specifically the vacuum
state. Accordingly, there is no need to add further
modes to purify them. Taking  this into
account, one can hence reduce the number
of environmental modes $\ell_{\text{pure}}$ that allows one to
represent $\Phi$ as in Eq.~(\ref{eq1}) in term of a {\em pure
state} $\hat{\rho}_E$ from $2n$ to
\begin{eqnarray}  \label{ellpure}
\ell_{\text{pure}} = n+ (n - r^\prime/2) = 2n -r^\prime/2\;,
\end{eqnarray}
i.e. we need the $n$ modes of $\alpha$ plus $n-r^\prime/2$
additional  modes of $\beta$ to purify those of $\alpha$ which
are not in a pure state already.
An easy way to characterize the parameter
$r^\prime$ is to observe that, according to Eq.~(\ref{Mequ1}), it
corresponds to the number of eigenvalues having modulus $1$ of the
matrix of $O\Sigma^\prime O^T$, i.e.,
\begin{eqnarray}
r^\prime &=& 2 n - \mbox{rank}[ \openone_{2n} - O \Sigma^\prime
(\Sigma^{\prime})^T O^T] = 2 n - \mbox{rank}[ \openone_{2n} -
\Sigma^\prime (\Sigma^{\prime})^T ] \nonumber \\&=& 2 n -
\mbox{rank}[ Y - \Sigma \, Y^{-1}\,  \Sigma^T ]
\label{defrprime}\,.
\end{eqnarray}
The explicit expressions for corresponding values of  $\gamma_E$ and
$s_2$ are given in Appendix~\ref{app:sols}. Here we notice that  for
$r^\prime = r =2n$ we get $\ell_{\text{pure}} =n$. This should
correspond to the channels~(\ref{minimalchannel}) of class (i)  for
which  one can construct a unitary dilation with pure input states.
Indeed, according to Eq.~(\ref{defrprime}),  when $r^\prime =2n$ the
matrix $Y  -\Sigma \, Y^{-1}\,  \Sigma^T$ must be zero, leading to
the identity~(\ref{minimalchannel}).

Taking into account that $r^\prime \leqslant r = \rank[\Sigma]$, a
further reduction in the number of modes $\ell$ can be obtained by
dropping the requirement of $\gamma_E$ being a  minimal
uncertainty covariance matrix. Indeed, an alternative unitary
representation~(\ref{eq1}) of $\Phi$ can be constructed with only
  \begin{eqnarray}
  \ell = n + (n- r/2) = 2n -r/2\,,
  \end{eqnarray}
environmental modes (see Appendix~\ref{app:sols2} for the explicit
solution).

The whole analysis can be finally generalized to  the BGCs of
class (iii), corresponding to channels that have non invertible matrices
$Y$. We have seen in fact that, in this case, the state
$\hat{\rho}_E$ which provides us the unitary dilation of Theorem 1
is constructed by replacing $Y$ with the strictly positive
operator $\bar{Y}$ of Eq.~(\ref{defbary}). Therefore for these
channels $\ell_{\text{pure}}$ of Eq.~(\ref{ellpure}) is defined by
Eq.~(\ref{defrprime}) with $Y$ replaced by $\bar{Y}$, i.e.
\begin{eqnarray}
r^\prime = 2 n - \mbox{rank}[ \bar{Y} - \Sigma \, \bar{Y}^{-1}\,
\Sigma^T ] \label{defrprimeiii}\,.
\end{eqnarray}
Taking $\bar{Y} := Y + (\openone_{2n} - \Pi)$ with $\Pi$ being the
projector on $\mbox{Supp}[Y]$ this gives,
\begin{eqnarray}
r^\prime &=&
2 n - \mbox{rank}[ \bar{Y} - \Sigma \, {Y}^{\ominus1}\,  \Sigma^T ]  =
2 n - \mbox{rank}[ {Y} - \Sigma \, {Y}^{\ominus1}\,  \Sigma^T ] -
\mbox{rank}[ \openone_{2n} - \Pi]
\nonumber \\
&=& k - \mbox{rank}[ {Y} - \Sigma \, {Y}^{\ominus1}\,  \Sigma^T ] \,,
\end{eqnarray}
where $k = \rank[Y] = \rank[\Pi]$, where ${Y}^{\ominus1}:= \Pi
\bar{Y}^{-1} \Pi$ denotes the Moore-Penrose inverse~\cite{HJ}  of
$Y$, and where we have
used the fact that $\mbox{Supp} [ \Sigma]
\subseteq \mbox{Supp}[Y]$. Remembering then  that for channels of
class (ii) $k=2n$ and $Y^{\ominus 1 } = Y^{-1}$ these results can
be summarized as follows:

\begin{theorem}[Dilations of BGCs involving
fewer additional modes]
Given  $\Phi$
a BGC  described by matrices $X$ and $Y$ satisfying
the conditions~(\ref{cptcond}) and characterized by the quantities
\begin{eqnarray} \label{IMPORTANTISSIMO}
    r=\mbox{rank}[ \Sigma]\,, \qquad
     r^\prime =\rank[Y] -
    \mbox{rank}[ Y - \Sigma Y^{\ominus1} \Sigma^T]\,.
\end{eqnarray}
Then it is possible to construct a unitary dilation~(\ref{eq1}) of
Stinespring form (i.e., involving a pure Gaussian state
$\hat{\rho}_E$) with  at most $\ell_{\text{pure}} = 2n
-r^\prime/2$ environmental modes. It is also
always possible to construct a unitary dilation~(\ref{eq1})
using $\ell = 2n -r/2$ environmental modes which
are prepared in a Gaussian, but not necessarily pure state.
\end{theorem}

It is worth stressing that, for channel of class (ii) and (iii),
the Theorem 2 only provides upper bounds for the minimal values of
$\ell$ and $\ell_{pure}$. Only in the generic
case (i) these bounds coincide with the real minima.

\subsection{Minimal noise channels}
\label{minimalnoise}

In a very analogous fashion to the extremal covariance matrices
corresponding to pure Gaussian states, one can introduce the
concept of a minimal noise channel. In this section we review the
concept of such minimal noise channels~\cite{Lindblad} and provide
criteria to characterize them. Given $X, Y\in \rr^{2n\times 2n}$
satisfying the inequality~(\ref{cptcond}), any other $Y^\prime = Y
+ \Delta Y$ with $\Delta Y \geqslant 0$ will also satisfy such
condition, i.e.,
\begin{eqnarray}
Y^\prime \geqslant Y \geqslant i ( \sigma_{2n} - X^T \sigma_{2n} X )
 \label{cptcondNN}\,.
\end{eqnarray}
Furthermore, due to the compositions rules~(\ref{NCOMP}), the BGC
$\Phi^\prime$ associated with the matrices $X, Y^\prime$ can be
described as the composition
\begin{eqnarray}
    \Phi^\prime = \Psi \circ \Phi
     \,, \label{mindec}
 \end{eqnarray}
between  the channel $\Phi$ associated with the matrices  $X, Y$,
and  the channel $\Psi$ described by the matrices
$X =\openone_n$
and $Y= \Delta Y$. The latter belongs to a special case  of BGC
that includes the so called   {\em additive classical noise
channels}~\cite{CGH,HW,EW} -- see
Sec.~\ref{sec:add} for details.

For any $X\in \rr^{2n\times 2n}$, one can then ask how much {\em
noise} $Y$ it is necessary to add in order to obtain a  map
satisfying the condition~(\ref{cptcond}). This gives rise to the
notion of {\it minimal noise}~\cite{Lindblad}, as the extremal
solutions $Y$ of Eq.~(\ref{cptcond}) for a given $X$. The
corresponding {\em minimal noise channels} are  the natural
analogue of the Gaussian pure state and allows one to represent
any other BGC as in Eq.~(\ref{mindec}) with a proper choice of the
additive classical noise map $\Psi$.

Let us start  considering  the case  of a generic channel
$\Phi^\prime$ of class (i) described by matrices $X$ and $Y^\prime$.
According to Theorem 1 it admits unitary dilation with $\ell =n$
modes described by some covariance matrix $\gamma_E^\prime$
satisfying the condition
\begin{eqnarray}
Y^\prime =  s_2 \, \gamma_E^\prime \, s_2^T
 \label{probb1NN} \,,\end{eqnarray}
for some proper $2n\times 2n$ real matrix $s_2$.
According to Eq.~(\ref{Ngammasimp11})
 $\gamma_E$  can be written as
 \begin{eqnarray}\label{gamam}
\gamma_E^\prime = \gamma_E  + \Delta \,,
\end{eqnarray}
with $\Delta\geqslant0$ and $\gamma_E$ minimal uncertainty state.
Therefore writing $Y= s_2 \gamma_Es_2^T$ and $\Delta Y = s_2 \Delta
s_2^T$ we can express $\Phi^\prime$ as in~(\ref{mindec}), where now
$\Phi$ is the BGC associated with the minimal noise environmental
state $\gamma_E$. Most importantly since the
decomposition~(\ref{gamam}) is optimal for $\gamma_E^\prime$, the
channel  $\Phi$ is an extremal solution of  Eq.~(\ref{cptcond}). We
stress that by construction $\Phi$ is still a channel of class (i):
in fact  it has the same $\Sigma$ as $\Phi^\prime$, while $Y$ is
still strictly positive since $\gamma_E >0$ and $s_2$ is invertible
-- see Eq.~(\ref{probb1NN}). We can then use the results of
Sec.~\ref{sec:red} to claim that $\Phi$ must satisfy the
equality~(\ref{minimalchannel}). This leads us  to establish three
equivalent necessary and  sufficient  conditions for minimal noise
channels of class (i):
\begin{eqnarray}
&(m_1)& \qquad  Y = \Sigma Y^{-1} \Sigma^T \label{cond1} \,, \\
&(m_2)& \qquad  \Det[Y] = \Det[\Sigma]  \label{cond2} \,, \\
&(m_3)& \qquad r= r^\prime   \label{cond3} \,,
\end{eqnarray}
with $r$ and $r^\prime$ as in Eq.~(\ref{IMPORTANTISSIMO}).
Since for class (i) we have that
$r=2n$, the minimal noise condition $m_3$
simply requires  the eigenvalues of the
matrix $\mu$ of Eq.~(\ref{Mequ1})
to be  equal to unity. Similarly, minimal noise
channels in case (ii) and (iii) can be characterized.

\begin{theorem}[Minimal noise condition]
A Gaussian bosonic channel characterized by the matrices $Y$ and
$X\in \rr^{2n\times 2n}$ is a minimal noise channel if and only if
\begin{equation}
    Y=\Sigma Y^{\ominus1} \Sigma^T,
\end{equation}
where, as throughout this work,
$\Sigma = \sigma_{2n} - X^T \sigma_{2n} X$.
\end{theorem}
{\em Proof:} The
complete positivity condition~(\ref{cptcond})
 of a generic  BGC
is a positive semi-definite constraint for the
symplectic form $\Sigma$, corresponding to
the constraint $\gamma - i\sigma_{2n}\geqslant 0$
in case of covariance matrices of
states of $n$ modes.
 In general, $r=\text{rank}[\Sigma]$ is not maximal, i.e., not equal to $2n$.
When identifying the minimal solutions of the inequality~(\ref{cptcond}), without loss of generality we can look for
the minimal solutions of
\begin{equation}\label{ms}
    Y' - i \Sigma'\geqslant 0,
\end{equation}
where here
\begin{eqnarray}
 \Sigma^\prime=      \left[
    \begin{array}{c|c}
       \begin{array}{c|c}
    \;\;0 \;\; &  \mu \\ \hline
    - \mu &  \;\;{0}\;\;
    \end{array} \\
    \hline
    &   {0}
     \end{array}
    \right] ,
    \end{eqnarray}
    with $\mu>0$ being diagonal of rank $r/2$
    ( here $Y' = O Y O^T$ and $\Sigma' = O\Sigma O^T$
    with $O\in {O}(2n, \rr)$ orthogonal).
The minimal solutions of inequality
(\ref{ms}) are then given by
$Y'= S S^T\oplus 0$, where $S$ is a $r\times r$
matrix satisfying
\begin{equation}
    S  \left[
       \begin{array}{cc}
    0  &  \mu \\
    - \mu & \;\; {0}\;\;
    \end{array}
    \right] S^T =  \left[
       \begin{array}{cc}
    0  &  \mu \\
    - \mu &\;\;  {0}\;\;
    \end{array}
    \right] ,
\end{equation}
so a symplectic matrix with respect to the
modified symplectic form, so
an element of
$\{M \in Gl(r,\rr): M=
(\mu^{1/2} \oplus\mu^{1/2})
S (\mu^{-1/2}\oplus \mu^{-1/2}),
S\in Sp(r,\rr)\}$.
From this, it follows that the minimal solutions of
(\ref{ms}) are  exactly given by the solutions
of $Y' = \Sigma' (Y')^{\ominus1} (\Sigma')^T$,
from which the statement of the theorem
follows.
$\blacksquare$
\newline

\subsection{Additive classical noise channel} \label{sec:add}

In this subsection we focus on the maps $\Psi$ which enter in the
decomposition~(\ref{mindec}). They are characterized by having $X=
\openone_{2n}$ and $Y\geqslant 0$. Note that with this choice the
condition~(\ref{cptcond}) is trivially satisfied. This is the
classical noise channel that has frequently been considered in the
literature (for a review, e.g., see Ref.\ \cite{EW}). For
completeness of the presentation, we briefly discuss this class of
multi-mode BGCs.

If the matrix $Y$ is strictly positive, the channel $\Psi$ is the
multi-mode generalization of the single mode additive classical
noise channel~\cite{CGH,HW,EW}. In the language of Ref.\
\cite{CGH}, these are the maps which have a canonical form $B_2$.
Indeed, one can show that  these maps are the  (Gaussian) unitary
equivalent to a collection of $n$ single mode additive classical
noise maps. To see this, let us apply symplectic transformations
($S_1$ and $S_2$) before and after the channel $\Psi$. Following
Eq.~(\ref{NXY}) this leads to $\{\openone_n, Y \} \ \mapsto \
\{S_1 S_2, \ \ S_2^T Y S_2\}$. Now, since $Y>0$, according to {\it
Williamson's theorem} \cite{Williamson}, we can find a $S_2\in
Sp(2n,\rr)$ such that $S_2^T Y S_2$ is diagonal
$\textrm{diag}(\lambda_1,\cdots,\lambda_n,\lambda_1,\cdots,\lambda_n)$
with $\lambda_i>0$. We can then take  $S_1=S_2^{-1}$ to have $S_1
S_2 = \openone_{2n}$. For $Y\geqslant 0$ but not $Y>0$, the maps
$\Psi$ that enter the decomposition Eq.~(\ref{mindec}) however
include also channels which are not unitarily equivalent to a
collection of $B_2$ maps. An explicit example of this situation is
constructed in Appendix \ref{ideal}.

\subsection{Canonical form for generic channels}
\label{sec:canonical}

Analogously to Refs.~\cite{CGH,SEW,HOLEVONEW}, any BGC  $\Phi$
described by the transformation Eq.~(\ref{Channel}) can be
simplified through unitarily equivalence by applying unitary
canonical transformations before and after the action of the
channel which induces transformations of the form~(\ref{NXY}).
Specifically, given  a $n$-mode Gaussian channel $\Phi$ described
by matrix $X$ and  $Y$  we can transform it into a new $n$-mode
Gaussian channel $\Phi_c$ described by the matrices
\begin{eqnarray}
X_c = S_1 X S_2 \,, \qquad  Y_c  =S_2^T Y S_2 \,,
\end{eqnarray}
with $S_{1,2}\in Sp(2n, \rr)$.
As already discussed in the introductory
sections, from an information theoretical
perspective $\Phi$ and $\Phi_c$ are equivalent in the
sense that, for instance, their unconstrained quantum
capacities coincide. We can then simplify
the analysis of the $n$-mode Gaussian channels by properly
choosing $S_1$ and $S_2$  to induce a parametrization of the
interaction part (i.e., $X$) of the evolution. The resulting
canonical form follows from the generalization of the Williamson
theorem~\cite{Williamson} presented in Ref.~\cite{WOLFLAST}.
According to this result,  for every non-singular
matrix $X\in Gl(2n,\rr)$, there
exist
matrices $S_{1,2} \in Sp(2n,\rr)$ such that
 \begin{eqnarray}
     \label{eq:normalform}
     X_c = S_1
    X S_2=\left[ %
    \begin{array}{cc}
      \openone_n & \,\,0\,\, \\
      \,\,0\,\, & J^T \\
    \end{array}%
    \right] \,,
 \end{eqnarray}
where $J^T$ is a $n \times n$ block-diagonal matrix in the real
Jordan form~\cite{HJ}. This can be developed a little further by
constructing a canonical decomposition for the symplectic matrix $S$
associated with the unitary dilation~(\ref{eq1}) of the channel.

For the sake of simplicity in the following we will focus on the
case of generic quantum channels $\Phi$ which have
non-singular $X\in Gl(2n,\rr)$
and belong to the class (i) of Theorem 1  (i.e.,
which have $r = \rank[ \Sigma ] =2n$). Under these
conditions $X$ must admit a
canonical decomposition of the form~(\ref{eq:normalform}) in which
all the eigenvalues of $J$ are different from $1$. In fact one has
\begin{eqnarray}
\Sigma = \sigma_{2n} - X^T \sigma_{2n} X = S_2^{-T} \, \left[
\sigma_{2n} - X_c^T \sigma_{2n} X_c \right] \, S_2^{-1} =
S_2^{-T}\,  \Sigma_c \, S_2^{-1}\,,
\end{eqnarray}
with $\Sigma_c$ being the skew-symmetric matrix associated with
the channel
 $\Phi_c$, i.e.,
\bea
 \Sigma_c:=\left[%
\begin{array}{cc}
{0} & \openone_n -J \\
 J^T - \openone_n & {0} \\
\end{array}%
\right]\,.
 \eea
Since $\rank[ \Sigma_c ] = \rank [\Sigma ] =2n$, it follows that
$J$ cannot have eigenvalues equal to $1$. Similarly, it is
not difficult to see that if
$X$ has a canonical form~(\ref{eq:normalform}) with all the
eigenvalues of $J$ being different from $1$, then $\Phi$ and
$\Phi_c$ are of class (i). However, a special case in which
$X=\openone_{2n}$ is investigated in details in Appendix
\ref{ideal}.

Consider then a unitary dilation~(\ref{eq1}) of the channel
$\Phi_c$ constructed with a not necessarily pure Gaussian state
$\hat{\rho}_E$ of $\ell = n$ environmental modes.
According to the above considerations,
such a dilation always exists. Let
 $S\in Sp(4n, \rr)$ be the
 $4n \times 4n$ real symplectic
transformation~(\ref{NmatrixS})  associated with the
corresponding unitary $\hat{U}$.
Assuming $s_1 = X_c^T$, an explicit expression for this
dilation can be
obtained by writing
\begin{equation}
    s_4 = \left[
    \begin{array}{cc}
      \openone_n & {0} \\
    {0} & J' \\
    \end{array}\right],\,\,
    s_{j} = \left[%
     \begin{array}{cc}  F_j & {0} \\
     {0} & G_j
    \end{array}%
    \right],
\end{equation}
where, for $j=2,3$,  $F_j$, $G_j$ are $n \times n$ real
matrices. Imposing Eqs.\ (\ref{NSN}), one obtains the following
relations
 \bea
&&J^T+ F_2 G_2^T = \openone_n \,, \qquad \qquad\qquad
 J'^T+F_3 G_3^T = \openone_n \,,  \nonumber \\
&& G_3^T+ F_2 J'^T = {0}  \,, \qquad  \qquad \qquad
G_2^T+ F_3 J^T = {0}  \,,
 \eea
whose solution gives an $S\in Sp(4n,\rr)$ of the form
 \bea
S = \left[%
\begin{array}{cc|cc}
  \openone_n & {0} & (\openone_n-J^T)G^{-T} & {0} \\
  {0} & J & {0} & G \\ \hline
  -G^T J^{-T} & {0} & \openone_n & {0}   \\
  {0} & G^{-1}J(J- \openone_n) & {0} & G^{-1} J G \\
\end{array}%
\right] \, , \label{Sprecanonical}
 \eea
with $G$ being an arbitrary matrix $G\in Gl(n,\rr)$.
As a consequence of this fact,
and because the eigenvalues of $J$ are assumed to be
different from $1$, $s_2$, $s_3$ and  $s_4$ are also non-singular.
This  is important because it shows that in choosing $S$ as in the
canonical form~(\ref{Sprecanonical}) we are not
restricting
generality: The
value of $s_2$ can always be absorbed into the
definition of the covariance matrix  $\gamma_E$ of $\hat{\rho}_E$
by writing
\begin{eqnarray}
\gamma_E  = s_2^{-1} \, Y_c \, s_2^{-T} \,,
\end{eqnarray}
(see also Appendix~\ref{sec:equivalent}). Taking this into account,
we can conclude that Eq.~(\ref{Sprecanonical}) provides an explicit
demonstration of Lemma 1 for channels of class (i) with non-singular
$X$.

Since $\Phi_c$ is fully determined  by $X_c$ and $Y_c$, the above
expressions show that the action of $\Phi_c$ on the input state
does not depend on the choice of $G$. As a matter of fact,   the
latter can be seen as  a Gaussian unitary operation $\hat{U}_G$
characterized by the $n$-modes symplectic transformation $Sp(2n,
\rr)$,
\begin{eqnarray}
\Delta_{G}=\left[%
\begin{array}{c|c}
  G^T & {0} \\ \hline
  {0} & G^{-1} \\
\end{array}%
\right]\,,
 \end{eqnarray}
applied to final state of the environment after the interaction with
the input, i.e., $\tilde{\Phi}_{G}=\hat{U}_{G} \tilde{\Phi}
\hat{U}_{G}^T$, where $\tilde{\Phi}$ is the {\it weak complementary
map} for $G =\openone_n$, and $\tilde{\Phi}_G$ is the weak
complementary map in presence of $G\neq \openone_n$ -- see the next
section for details. Since the relevant properties of a channel
(e.g., weak degradability~\cite{CG,CGH}) do not depend on local
unitary transformations to the input/output states,  without loss of
generality, we can consider $G=-J$ and the canonical form for $S\in
Sp(4n,\rr)$ assumes the following simple expression
 \bea
 \label{canonS}
 S = \left[%
\begin{array}{cc|cc}
  \openone_n & \;\;\;\;{0}\;\;\;\; & \openone_n -J^{-T} & {0} \\
  \;\;\;\;{0} \;\;\;\;& J & {0} & -J \\ \hline
  \openone_n & {0} & \openone_n & {0}   \\
  {0} & \openone_n -J & {0} & J  \\
\end{array}%
\right] \, .
 \eea
The possibility of constructing different, but unitarily
equivalent, canonical forms for $S$ is discussed  in
Appendix~\ref{sec:equivalent}.

\section{Weak degradability} \label{sec:weak}

Among other properties, the unitary dilations introduced in Section
\ref{sec:unitary} are useful  to define {\it complementary} or {\it
weak complementary} channels of a given BGC $\Phi$. These are
defined as the  CPT  map ${\tilde{\Phi}}$  which describes the
evolution of the environment  under the influence of the physical
operation describing the channel \cite{CG,CGH}, i.e.,
\begin{eqnarray} \label{defwconj}
    \tilde{\Phi}(\hat{\rho}) := \mbox{Tr}_S[ \hat{U} (
    \hat{\rho} \otimes
    \hat{\rho}_E)\hat{U}^\dag] \label{Nphicmpl} \,,
\end{eqnarray}
where $\hat{\rho}$, $\hat{\rho}_E$ and $\hat{U}$ are defined as in
Eq.~(\ref{eq1}), but the partial trace is now  taken over the
system modes.

Specifically, if the state $\hat{\rho}_E$ we employed in
constructing the unitary dilation of  Eq.~(\ref{eq1}) is pure, then
the map $\tilde{\Phi}$ is said to be the {\em complementary} of
$\Phi$ and, up to partial isometry, it is
unique~\cite{DEVSHOR,HOLEVO11,KINGMATS,LIND1}. Otherwise it is
called {\em weak complementary}~\cite{CG,CGH}. Since in
Eq.~(\ref{eq1}) the state $\hat{\rho}_E$ is Gaussian and $\hat{U}$
is a unitary Gaussian transformation, one can verify that
$\tilde{\Phi}$ is also BGC\footnote{In general however, it will not
map the $n$ input modes into $n$ output modes. Instead  it will
transform them into $\ell$ modes, with $\ell$ being the number of
modes  assumed in the unitary dilation~(\ref{eq1}).}. Expressing the
Gaussian unitary transformation $\hat{U}$ in terms of its symplectic
matrix $S$ of Eq.~(\ref{NmatrixS}) the action of $\tilde{\Phi}$ is fully
characterized by the following mapping of the covariance matrices
$\gamma$  of $\hat{\rho}$, i.e.,
     \bea
     \tilde{\Phi}: \ \gamma \longmapsto \ s_3
    \gamma s_3^T + s_4 \gamma_E s_4^T \,, \eea
which is counterpart of the transformations~(\ref{NChannel}) and
(\ref{dil}) that characterize $\Phi$. The channel $\tilde{\Phi}$ is
then described by the matrices $\tilde{X}=s_3^T$ and $\tilde{Y}=s_4
\gamma_E s_4^T$ which, according to the symplectic
properties~(\ref{NSN}), satisfy the condition
\begin{equation}\label{cptcondconj}
\tilde{Y} \geqslant i  \tilde{\Sigma}\,   \qquad \mbox{with} \quad
\tilde{\Sigma} :=  \sigma_{2\ell}^E  - \tilde{X}^{T}
\sigma_{2n} \tilde{X}  \,.
\end{equation}

The relations between  ${\Phi}$ and its weak complementary
$\tilde{\Phi}$ contain useful information about the channel $\Phi$
itself. In particular we say that the channel ${\Phi}$ is {\it
weakly degradable} (WD) while $\tilde{\Phi}$ is {\it
anti-degradable} (AD), if there exists a CPT map ${\cal T}$ which,
for all inputs $\hat{\rho}$, allows one to recover
$\tilde{\Phi}(\hat{\rho})$ by acting on the output state
$\Phi(\hat{\rho})$, i.e.
\begin{eqnarray}
{\cal T} \circ \Phi =
\tilde{\Phi}\,.\label{weakdegdef}
\end{eqnarray}
Similarly, one says that $\Phi$ is AD and $\tilde{\Phi}$ is WD if
there exists a CPT map $\bar{\cal T}$ such that
\begin{eqnarray}
\bar{\cal T} \circ \tilde{\Phi} = {\Phi}\,.\label{weakantidegdef}
\end{eqnarray}

Weak degradability~\cite{CG,CGH} is a property of quantum channels
${\Phi}$ generalizing the {\it degradability property} introduced in
Ref.~\cite{DEVSHOR}. The relevance of weak-degradability analysis
stems from the fact that it allows one to simplify the quantum
capacity scenario. Indeed, it is known that AD  channels have zero
quantum capacity~\cite{CG,CGH}, while WD channels with
$\hat{\rho}_E$ pure are degradable and thus admits a single letter
expression for this quantity~\cite{DEVSHOR}. A complete
weak-degradability analysis of single mode bosonic Gaussian channels
has been provided in Ref.\ \cite{CG,CGH}. Here we generalize some of
these results to $n >1$.

\subsection{A criterion for weak degradability} \label{sec:criterion}

In this section we review a general criterion for degradability of
BGCs which was introduced in Ref.~\cite{WGG}, adapting it to include
also weak degradability.
 Before entering the details of our derivation, however, it
is worth noticing that generic multi-mode Gaussian channels are
neither WD nor AD. Consider in fact a WD single-mode Gaussian
channel $\Phi$ having no zero quantum capacity $Q>0$ (e.g., a
beam-splitter channel with transmissivity $> 1/2 $). Define then
the two mode channel ${\Phi}\otimes \tilde{\Phi}$ with
$\tilde{\Phi}$ being its weak complementary defined
in~\cite{CG,CGH}. This  is Gaussian since both $\Phi$ and
$\tilde{\Phi}$ are Gaussian. The claim is that ${\Phi}\otimes
\tilde{\Phi}$ is neither WD nor AD. Indeed, its weak complementary
can be identified with the map $\tilde{\Phi}\otimes {\Phi}$.
Consequently, since ${\Phi}\otimes \tilde{\Phi}$ and
$\tilde{\Phi}\otimes {\Phi}$ differ by a permutation, they must
have the same quantum capacity $Q^\prime$. Therefore if one of the
two is WD than {\em both} of them must also be AD. In this case
$Q^\prime$ should be zero which is clearly not possible given that
$Q^\prime \geqslant Q$. In fact, one can use $\Phi\otimes
\tilde{\Phi}$ to reliably transfer quantum information by encoding
it into the inputs of $\Phi$. In this respect the possibility of
classifying (almost) all single-mode Gaussian maps in terms of
weak degradability property turns to be rather a remarkable
property. We now turn to investigating the weak degradability
properties of multi-mode bosonic Gaussian channels deriving a
criterion that will be applied in Sec.~\ref{wd2} for studying in
details the two-mode channel case.

Consider a $n$-mode bosonic Gaussian channel $\Phi$ characterized
by the unitary dilation~(\ref{eq1}) and its weak complementary
$\tilde{\Phi}$~(\ref{Nphicmpl}). Let $\{X,Y\}$,
$\{\tilde{X},\tilde{Y}\}$  be the matrices which define such
channels. For the sake of simplicity we will assume $X$ and
$\tilde{X}$ to be non-singular, with $X,\tilde X\in Gl(2n,\rr)$.
Examples of such maps are for instance the channels of class (i)
with $X$ non-singular described in Sec.~\ref{sec:canonical}.
Adopting in fact the canonical form~(\ref{canonS}) for $S$ we have
that
\begin{equation}
    X = \left[ %
    \begin{array}{cc}
      \openone_n &0\\
      0& J^T \\
    \end{array}%
    \right] ,\qquad
    \tilde{X} = \left[ %
    \begin{array}{cc}
      \openone_n & 0 \\
      \;\;\;0\;\;\;& \openone_n - J^T\\
    \end{array}%
    \right]
\end{equation}
with all the eigenvalues of $J$ being different from  $1$.

Suppose now that $\Phi$ is weakly degradable with $\cal T$ being
the connecting CPT map which satisfies the weak degradability
condition~(\ref{weakdegdef}). As in Refs.\
\cite{CG,CGH} we will focus on
the case in which ${\cal T}$ is BGC and described by matrices
$\{X_{\cal T},Y_{\cal T}\}$. Under these hypothesis the
identity~(\ref{weakdegdef}) can be simplified by using the
composition rules for BGCs given in Eq.~(\ref{NCOMP}).
Accordingly, one must have
\begin{eqnarray}
X_{\cal
T}&=&X^{-1} \tilde{X} \,, \nonumber \\
Y_{\cal T}&=& \tilde{Y} - X_{\cal T}^T Y
X_{\cal T}\,. \label{afaf}
\end{eqnarray}
These definitions must be compatible with the requirement that
$\cal T$ should be a CPT map which transforms the $n$ system modes
into the $\ell$ environmental modes, i.e.,
\begin{eqnarray}
Y_{\cal T} \geqslant  i\left( \sigma_{2\ell}^E - X_{\cal T}^T
\sigma_{2n} X_{\cal T} \right) \,.
\end{eqnarray}
Combining the expressions above, one finds the following
weak-degradability condition for $n$-mode bosonic Gaussian
channels~\cite{WGG}, i.e.
 \bea  \label{nmode-deg}
\tilde{Y} - \tilde{X}^T X^{-T} (Y+i \sigma_{2n}) X^{-1} \tilde{X} + i
\sigma_{2\ell}^E \geqslant 0 \, .
 \eea
In order to obtain the anti-degradability
condition~(\ref{weakantidegdef}), it is sufficient to swap
$\{X,Y\}$ with $\{\tilde{X},\tilde{Y}\}$ and the system commutation
matrix $\sigma_{2n}$ with $\sigma_{2\ell}^E$, in
Eq.~(\ref{nmode-deg}), i.e.,
\bea
Y - X^T \tilde{X}^{-T} (\tilde{Y}+i
    \sigma_{2\ell}^E) \tilde{X}^{-1} X + i \sigma_{2n} \geqslant 0 \,.
    \label{antideg}
 \eea
Equations~(\ref{nmode-deg}) and (\ref{antideg}) are strictly
related. Indeed since
 \bea
    &&Y - X^T \tilde{X}^{-T} (\tilde{Y}+i \sigma_{2\ell}^E) \tilde{X}^{-1} X
    + i \sigma_{2n} \\
    && \qquad \qquad \quad = -X^T \tilde{X}^{-T} \left(\tilde{Y} -
    \tilde{X}^T X^{-T} (Y+i \sigma_{2n}) X^{-1} \tilde{X} + i
    \sigma_{2\ell}^E \right) \tilde{X}^{-1} X \nonumber \,,
 \eea
equation (\ref{antideg}) corresponds to reverse the sign of the
inequality~(\ref{nmode-deg}), i.e.
 \bea
\tilde{Y} - \tilde{X}^T X^{-T} (Y+i \sigma_{2n}) X^{-1} \tilde{X} + i
\sigma_{2\ell}^E \leqslant 0 \, .
 \eea
Hence to determine if $\Phi$ is a weakly degradable or
anti-degradable channel, it is then sufficient to study the
positivity of the Hermitian matrix
\begin{eqnarray}
W := \tilde{Y} - \tilde{X}^T X^{-T} (Y+i \sigma_{2n}) X^{-1} \tilde{X} + i
\sigma_{2\ell}^E \label{thematrix}\,.
\end{eqnarray}
In the case in which $\ell =n$ this can be
simplified by reminding that an Hermitian $2n \times 2n$ matrix $W$
partitioned
as
\begin{equation}
W=\left[%
\begin{array}{cc}
  W_1 & W_2 \\
  W_2^{\dagger} & W_3 \\
\end{array}%
\right]
\end{equation}
with $W_i$ being $n \times n$ matrices
 is semi-positive definite if and only if
 \bea
W_1 \geqslant 0 \, \, \, \textrm{and} \, \, \, W_3 -W_2^{\dagger}
W_1^{-1} W_2 \geqslant 0 \, , \label{conditE}
 \eea
 the right hand side being the Schur complement of $W$
 (see, e.g., page 472 in Ref.\ \cite{HJ}).
Using this result and the canonical form~(\ref{canonS}),
Eq.~(\ref{nmode-deg}) can be written as in Eq.~(\ref{conditE})
with
 \bea
    \label{Nsimplif}
    W_1 &=&(\openone_n-J^{-T})^{-1} Y_1 (\openone_n -J^{-1})^{-1}-Y_1 \\
    W_2 &=& i(J^{-T}-2 \openone_n)-Y_2 (J^{-T}-\openone_n)-(\openone_n
    -J^{-T})^{-1} Y_2 \nonumber \\
    W_3 &=& Y_3-(J^{-1}-\openone_n) Y_3 (J^{-T}-\openone_n ) \nonumber \, ,
     \eea
and
\begin{equation}
    Y=\left[%
    \begin{array}{cc}
      Y_1 & Y_2 \\
      Y_2^{T} & Y_3 \\
    \end{array}%
    \right].
\end{equation}
For the anti-degradability condition~(\ref{antideg})
simply replace $[\geqslant]$ with $[\leqslant]$ in
Eq.~(\ref{conditE}).

\section{Two-mode bosonic Gaussian channels}
\label{twomode}

Here we consider a particular case of $n$-mode bosonic Gaussian
channel analysis above, namely, the case of $n=2$.
This is by no means such a special case as one might
at first be tempted to think since
 any $n$-mode
channel can always be reduced to single-mode and two-mode parts~\cite{WOLFLAST}. For two-mode channels
the interaction
part and the noise term of a generic two-mode bosonic Gaussian
channel, $X$ and $Y$, respectively, are $4 \times 4$ real
matrices. Particularly, we will focus on two-mode channels $\Phi$
which have non-singular $X$ and belong to the class (i) of Theorem 1 (i.e., which
have $r = \rank[ \Sigma ] =4$), like in Sec. \ref{sec:canonical}.
These maps can be grasped in terms of a unitary dilation of the
form (\ref{canonS}) coupling the two system bosonic modes with two
additional (environmental) modes, where $J$ is a $2 \times 2$ real
Jordan block. In order to characterize this large class of
two-mode BGCs, one has to examine only three possible forms of
$J$:
\begin{itemize}
\item   Class A: This corresponds to taking a diagonalizable Jordan
block, that is,
 \bea
 \label{J0}
J := J_0 =\left[%
\begin{array}{cc}
  a & 0 \\
  0 & b \\
\end{array}%
\right] \, .
 \eea
where $a$ and $b$ are real nonzero numbers. It represents the
trivial case of a two-mode bosonic Gaussian channel, whose
interaction term does not couple the two modes.
Actually, we call it of class $A_1$ if $a \neq b$ and of class
$A_2$ otherwise.

\item Class B: This is to take $J$ as a non-diagonalizable matrix
with a nonzero real eigenvalue $a$ with double algebraic
multiplicity (but with geometric multiplicity equal to one), i.e.
 \bea
 \label{J1}
J := J_1 =\left[%
\begin{array}{cc}
  a & 1 \\
  0 & a \\
\end{array}%
\right] \, .
 \eea
In this case the Jordan block is called defective \cite{HJ}. Here,
a noisy interaction between the bosonic system and the
environment, coupling the two system modes, is switched on.

\item Class C: Here the real Jordan block
$J$ has complex eigenvalues, i.e.
 \bea
 \label{J2}
J := J_2 =\left[%
\begin{array}{cc}
  a & b \\
  -b & a \\
\end{array}%
\right] \, ,
 \eea
with $b \neq 0$; the eigenvalues of $J$ are $a \pm i b$.  Again,
the two system modes are coupled by the noisy interaction with the
environment through the presence of the element $b$.
\end{itemize}

In order to explicit the form of $Y=s_2 \gamma_E s_2^T$, with
$s_2$ being defined as in Eq.\ (\ref{canonS}), we consider a
generic two-mode covariance matrix in the so-called standard form
\cite{LAURAT} for the environmental initial covariance
matrix $\gamma_E$, i.e.
 \bea
 \gamma_E=\left[%
\begin{array}{cc}
  \Gamma_1 & 0 \\
  0 & \Gamma_2 \\
\end{array}%
\right] \, , \label{gamma2sta}
 \eea
where
 \bea
\Gamma_{1,2}
:=\left[%
\begin{array}{cc}
  x & z_{-,+} \\
  z_{-,+} & y \\
\end{array}%
\right] \, ,
 \eea
and $x$, $y$, $z_{+,-}$ are real number satisfying $x + y \geqslant 0$,
$x y - z_{-}^2 \geqslant 1$ and $x^2 y^2 -y^2 - x^2 + (z_{-} z_{+}-1)^2
- x y (z_{-}^2 + z_{+}^2) \geqslant 0$ because of the uncertainty
principle. More generally, one can apply a generic
two-mode (symplectic) squeezing operator $V(\epsilon)$ to the
environmental input state, i.e.,
 \begin{equation}
 \gamma_E'=V(\epsilon) \gamma_E
V(\epsilon)^T \label{gamma2mod}
 \end{equation}
where
\begin{equation}
    V(\epsilon)=\left[%
    \begin{array}{cc}
      R^{-T} & 0 \\
      0 & R \\
    \end{array}%
    \right],\,\,R=\left[%
    \begin{array}{cc}
      c + h s & -q s \\
      -q s & c - h s \\
    \end{array}%
    \right],
\end{equation}
and $c=\cosh(2r)$, $s=\sinh(2r)$, $h=\cos(2\phi)$,
$q=\sin(2 \phi)$ and $\epsilon = r e^{2 i \phi}$ being the
squeezing parameter \cite{LAURAT}.
Finally, it is interesting to study how the canonical forms of
two-mode BGCs compose under the product. A simple calculation
shows that the following rules apply

%%%%%%%%%%%%%%%%%%%%%%%%%%%%%%%%%%%%%%%%%
\begin{eqnarray}
\begin{array}{c|ccc}
 \circ     & A & B & C\\ \hline
A & A & A_1/B & A_1/B/C \\
B & A_1/B & A_2/B & A_1/B/C \\
C & A_1/B/C & A_1/B/C & A/C
\end{array} \nonumber .
\end{eqnarray}
%%%%%%%%%%%%%%%%%%%%%%%%%%%%%%%%%%%%%%%%%

\
\newline
\

In this table, for instance, the element on row 1 and column 1
represents the class (i.e., $A$) associated to the composition of
two channels of the same class $A$. Note that the canonical form
of the products with a ``coupled'' channel (i.e., with $B$ or $C$)
is often not uniquely defined. For instance, composing two class
$B$ channels characterized by the matrices
\begin{equation}
    ({J_1})_i=\left[%
    \begin{array}{cc}
      a_i & 1 \\
      0 & a_i \\
    \end{array}\right]
\end{equation}
with $i=1, 2$, will give us either a class $A_2$ channel (if
$a_1+a_2=0$) or a class $B$ channel (if $a_1+a_2 \neq 0$).
Composition rules analogous to those reported above have been
analyzed in details for the one-mode case in Ref. \cite{CGH}. In
the following we will study the weak-degradability properties of
these three classes of two-mode Gaussian channels.

\subsection{Weak-degradability properties}
\label{wd2}

The weak-degradability conditions in Eqs. (\ref{conditE}) become
 \bea
\Gamma_1-(\openone_2-J^{-T}) \Gamma_1 (\openone_2-J^{-1}) \geqslant 0
 \eea
and
 \bea
 &&J \Gamma_2 J^{T} -(\openone_2-J)\Gamma_2 (\openone_2-J^{T})
\\
&& \qquad -(J^{-1}-2 \openone_2) \left[ \Gamma_1-(\openone_2-J^{-T}) \Gamma_1
(\openone_2-J^{-1})\right]^{-1} (J^{-T}-2 \openone_2) \geqslant 0 \,. \nonumber
 \eea
In the same way, the anti-degradability is obtained when both these
quantities are non-positive. As concerns the environmental initial
state of the unitary dilation, one can consider a generic two-mode
state as in Eq.\ (\ref{gamma2mod}). On one hand, we find that, if
$[J,R]=0$, this two-mode squeezing transformation $V(\epsilon)$
can be simply ``absorbed'' in local symplectic operations to the
output states and then it does not affect the weak-degradability
properties. On the other hand, if $[J,R] \neq 0$, we find
numerically that the introduction of correlations between the two
environmental modes contrasts with the presence of (anti-)
weak-degradability features. Therefore, one can consider the
particular case in which the environment is initially in a
state with a symmetric covariance matrix
$\gamma_E$ as in Eq.\ (\ref{gamma2sta}) with
$x=y=2N+1$ and $z_{-}=z_{+}=0$ where $N \geqslant 0$. In this case
$\gamma_E=(2 N + 1) \openone_2$ corresponds to a thermal state of
two uncoupled environmental modes with the same photon average
number $N$ and it is possible to see the results above easily
through analytical details. In fact, we study analytically the
positivity condition in Eq.\ (\ref{nmode-deg}) in the three
possible forms of the real Jordan block $J_i$.

In the uncoupled case $J_0$ as in Eq.\ (\ref{J0}), substituting in
Eq.\ (\ref{nmode-deg}), we find that these two-mode bosonic
Gaussian channels are WD if $a, b \geqslant 1/2$ and AD for $a, b \leq
1/2$ (any $N \geqslant 0$). In other words, in the case of two
uncoupled modes, the weak-degradability properties
can be derived from the results for one-mode bosonic Gaussian
channels: tensoring two WD (AD) one-mode Gaussian channels with WD
(AD) one-mode Gaussian channels yield two-mode Gaussian channels
which are WD (AD).

\begin{figure}[h!]
\begin{center}
\includegraphics[scale=.8]{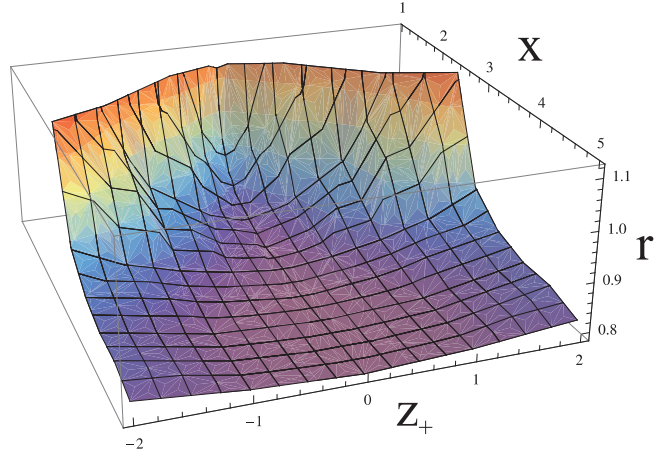}
\caption{Relation between the parameters $x$, $z_{+}$ and the
minimum value of $r$ in the initial environmental state such that
the two-mode channel with $X=\openone_2 \oplus J_1$ reduces to the
decoupled case $X'=\openone_2 \oplus J_0$ with the same interaction
parameter $a$ for the two system modes.} \label{fig3}
\end{center}
\end{figure}

\begin{figure}[h!]
\begin{center}
\includegraphics[scale=1.2]{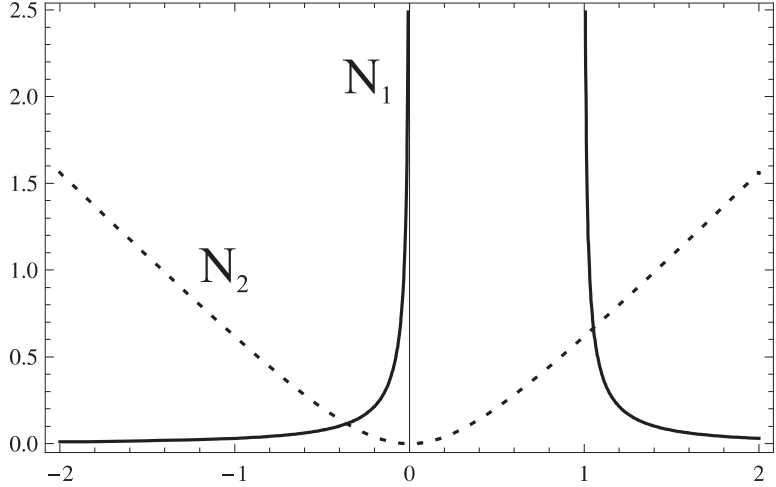}
\caption{In continuous line we report $N_1$ as function of $a$ in
the case of $J_1$. For $N \geqslant N_1$ the map is WD if $a
> 1$ and AD if $a <0 $. In dashed line we plot $N_2$
as function of $b$ when $a=0$ in the case of $J_2$. For $ N \geq
N_2$ the channel is WD (AD) if $a > 1/2$ ($a<1/2$).} \label{fig1}
\end{center}
\end{figure}

In the case of defective $J$, i.e., $J_1$ as in Eq.\ (\ref{J1}),
corresponding to noisy interaction coupling the two system modes,
substituting in Eq.\ (\ref{nmode-deg}), we find that, on one hand,
these two-mode bosonic Gaussian channels are WD if $a
> 1$ and
 \bea
 N \geqslant N_1 := \frac{1}{2}\left[-1+\frac{1}{2} \frac{|2 a -
1|}{\sqrt{a(a-1)}} \right] \, . \label{N1}
 \eea
On the other hand, it is AD if $a < 0$ and $N \geqslant N_1$ (see Fig.
\ref{fig1}). Note that the defective Jordan blocks are not usually
stable with respect to perturbations \cite{WOLFLAST}. Indeed, we
find numerically that, applying proper two-mode squeezing
transformations to the environmental input, these
weak-degradability conditions reduce to the decoupled case ones.
In Fig. \ref{fig3} we consider, for simplicity, a
symmetric environmental initial state $\gamma_E'$ as in Eq.
(\ref{gamma2mod}) with $x=y$, $z_{-}=0$ and $\epsilon = r$, and we
plot the relation between $x$, $z_{+}$ and the minimum value of
$r$ such that $J := J_1$ reduces to $J := J_0$
corresponding to the decoupled case. One realizes that a squeezing
parameter $r$ close to $1$ is enough to decouple the two modes
representing the system, carrying quantum information. Moreover,
let us point out that this squeezing threshold (r) increases
slightly with the presence of correlations ($z_{+}$) while
decreases when increasing the level of noise ($x$) in the initial
environmental state $\gamma_E'$.

%%%%%%%%%%%%%%%%%%%%%%%%%%%%%%%%%%%%%%%%%%%%%%%%%%%
%%%%%%%%%%%%%%%%%%%%%%%%%%%%%%%%%%%%%%%%%%%%%%%%

%%%%%%%%%%%%%%%%%%%%%%%%%%%%%%%%%%%%%%%%%%%%%%%%%%%
%%%%%%%%%%%%%%%%%%%%%%%%%%%%%%%%%%%%%%%%%%%%%%%%

Finally, in the case of real Jordan block with complex
eigenvalues, i.e., $J_2$ as in Eq.\ (\ref{J2}), the corresponding
two-mode bosonic Gaussian channels are WD if $a > 1/2$ and
 \bea
    N \geqslant N_2 := \frac{1}{2}
    \left[-1+\left(1+\frac{4 b^2}{(1-2 a)^2}\right)^{1/2}
\ \right] \, . \label{N2}
 \eea
while they are AD if $a < 1/2$ and $N \geqslant N_2$ (see Fig.
\ref{fig1}). In both of these cases (real and complex
eigenvalues), in which the interaction term couples the two
bosonic modes, there is the (apparently) counter-intuitive fact
that above a certain environmental noise threshold the
weak-degradability features appear, while for one-mode bosonic
Gaussian channels they do not depend on the initial state of the
environment. Actually, one would expect at most that, when the
level of the environmental noise increases, the coherence
progressively decreases until to be destroyed. It would mean that
it becomes more and more difficult to recover the environment
(system) output from the system (environment) output after the
noisy evolution. However, the things go the other way around when
multi-mode bosonic Gaussian channels are considered.

\subsection{Channels with zero quantum capacity}

Analogously to Ref. \cite{CGH} where the one-mode case is
investigated, one can enlarge (other than the AD maps) the class of
two-mode BGCs with $Q=0$, composing a generic channel with an AD
one. First of all, consider a channel $\Phi$ as in Section
\ref{minimalnoise}, but being AD (not necessarily minimal noise),
then the maps $\Phi'$, defined in Eq.\ (\ref{mindec}), have zero
quantum capacity, i.e., they cannot be used to transfer quantum
information. For instance, one can choose $\gamma_E=(2 N_c +1)
\openone_n$, i.e., the environmental initial state of the map $\Phi$
is a multi-mode thermal state with $N_c$ being the average photon
number for each mode, such that $\Phi$ is AD or simply with zero
capacity; therefore, for any $\gamma_E' \geqslant \gamma_E=(2 N_c+1)
\openone_n$, as in Eq.\ (\ref{gamam}), the map $\Phi'$ of Eq.
(\ref{mindec}) has $Q=0$. Particularly for $n=2$, using these observations and
choosing $N_c$ equal to either $N_1$ (and $a<0$) or $N_2$ (and
$a<1/2$) as in Eqs. (\ref{N1}) and (\ref{N2}), one obtains that for
$X=\openone_2 \oplus J_{1,2}$ and $Y'=s_2 \gamma_E' s_2^T$ [with
$s_2$ as in Eq.~(\ref{canonS})] the resulting channel $\Phi'$ has
always zero capacity. In this way, one extends considerably the set
of two-modes maps with zero capacity, other than the very particular
cases of two-mode environmental thermal states studied above and
shown in Fig. \ref{fig1}. For instance, two-mode squeezing can be
applied to the thermal state $\gamma_E$ including not only states
with $N > N_c$ but also with not trivial two-mode correlations such
that $\gamma_E' \geqslant (2 N_c+1) \openone_2$. Therefore, just
considering this last simple inequality one includes so a larger set
of maps that have zero quantum capacity.

\begin{figure}[h!]
\begin{center}
\includegraphics[scale=1.2]{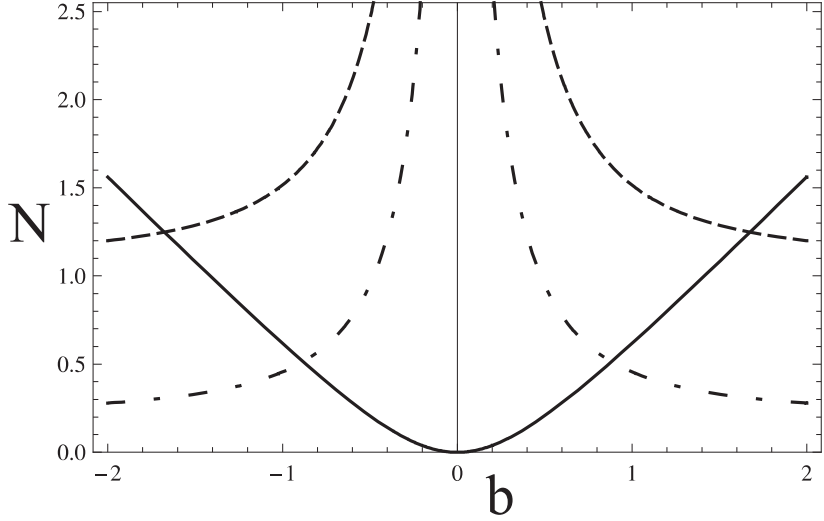}
\caption{The continuous line depicts
plot $N_2$ as in Eq.\ (\ref{N2})
versus $b$, with $a=1$ in $J_2$ of Eq.\ (\ref{J2}). For $ N \geq
N_2$ the channel is  WD (AD) if $a > 1/2$ ($a<1/2$). The dashed
line refers to the bound in Eq.~(\ref{bound}), while the
dashed-dot line to the one in Eq.~(\ref{bound1}); above these
bounds the class $C$ map is WD but with $Q=0$. Note that
Eq.~(\ref{bound1}) is an improvement with respect to the
constraint of Eq.~(\ref{bound}). Similar bounds can be obtained in
the case $a<1/2$, enlarging the group of AD maps with other
channels with $Q=0$.} \label{fig2}
\end{center}
\end{figure}

Moreover, we observe that, according to composition rules above,
the combination ${\Phi} = {\Phi}_{II} \circ {\Phi}_I$ of two
channels ${\Phi}_I$ and ${\Phi}_{II}$ of class $A_2$ and $C$,
respectively, with Jordan blocks $J_I$ as in Eq.\ (\ref{J0}) with
$a_I=b_I$ and $J_{II}$ as in Eq.\ (\ref{J2}) with $a_{II}$ and
$b_{II}\neq 0$, gives $J=a_I J_{II}$ which is in the class $C$.
Now, since we have $N_1 \geqslant 0$, $N_2 \geqslant 0$
and assuming $a_I \leqslant 1/2$, the channel ${\Phi}_I$ is AD and the
resulting channel $\Phi$ must have $Q=0$. Varying the parameters but
keeping the product $a_I a_{II}=a$ and $a_I b_{II}=b$ fixed, the
parameter $N$ can assume any value satisfying the inequality
\begin{eqnarray}
N  &\geqslant& \frac{1}{4}
\left[\left({\frac{5(1-4a+8a^2+8b^2)}{b^2+(a-1)^2}}\right)^{1/2} -2 \right]
\label{bound} \,.
\end{eqnarray}
Note that $a_I$ has been chosen equal to $1/2$ and ${\Phi}_I$
corresponds to two uncoupled beam-splitter maps with transmissivity
$1/2$. We can therefore conclude that all channels of the form $C$
with $N$ as in Eq.~(\ref{bound}) have zero quantum capacity -- see
Fig.~\ref{fig2}.

Consider now the composition ${\Phi} = {\Phi}_{II} \circ {\Phi}_I$
of two channels ${\Phi}_I$ and ${\Phi}_{II}$ of class $C$ and
$A_2$ (i.e., in the opposite order with respect to
above), respectively, with
Jordan blocks $J_I$ as in Eq.\ (\ref{J2}) with $a_I$ and $b_I\neq 0$ and
$J_{II}$ as in Eq.\ (\ref{J0}) with $a_{II}=b_{II}$, giving
$J=a_{II} J_I$ which is in the class $C$. As
before, since we have $N_1 \geqslant 0$, $N_2 \geqslant 0$ and assuming
again $a_{II} \leqslant 1/2$, the channel ${\Phi}_2$ is AD and the
resulting channel has $Q=0$. Varying the parameters but keeping
the product $a_I a_{II} = a$ and $b_I a_{II} = b$ fixed, the
parameter $N$ can assume any value satisfying the inequality
\begin{eqnarray}
N  &\geqslant& \frac{1}{4} \left[\left({\frac{(1+4a^2+4b^2)(1-4 a +
8 a^2+8 b^2)}{4 (b^2+(a-1)^2)(a^2+b^2)}}\right)^{1/2} -2 \right] \, ,
\label{bound1}
\end{eqnarray}
where again $a_{II}$ is chosen equal to $1/2$. Again we can conclude
that all class $C$ channels with $N$ as in Eq.~(\ref{bound1}) have
zero quantum capacity. However, notice that the constraint in Eq.\
(\ref{bound1}) is an improvement with respect to the constraint of
Eq.~(\ref{bound}) -- see Fig.~\ref{fig2}.

%%%%%%%%%%%%%%%%%%%%%%%%%%%%%%%%%%%%%%%%%%%%%%%
%%%%%%%%%%%%%%%%%%%%%%%%%%%%%%%%%%%%%%%%%%%%%%%%

\section{Conclusions}

In this work, we have presented a complete analysis of generic
multi-mode Gaussian channels by proving a unitary dilation theorem
and by finding their canonical form. This is a simple form that can
be achieved for any Gaussian quantum channel, as a convenient
starting point for various considerations. For instance, it allows
us to simplify the analysis of the weak-degradability properties of
multi-mode bosonic Gaussian channels. Minimal output entropies, or
quantum and classical information capacities and other difficult
questions might be tackled using the canonical form of multi-mode
Gaussian channels shown in this paper. Here, we investigated in
details the two-mode scenario that is relevant since any $n$-mode
channel can always be reduced to single-mode and two-mode parts
\cite{WOLFLAST}. Furthermore, the results of this paper could play a
basic role in characterizing the efficiency of continuous-variables
quantum information processing, quantum communication and quantum
key distribution protocols.

\section{Acknowledgements}

F.C.\ and V.G.\ thank the Quantum Information research program of
Centro di Ricerca Matematica Ennio De Giorgi of Scuola Normale
Superiore for financial support. J.E.\ acknowledges the EPSRC, the
QIP-IRC, the EU (QAP, COMPAS),
Microsoft Research, and the EURYI Award
Scheme for financial support. A.S.H.\ acknowledges support of RFBR
grant 06-01-00164-a and the RAS program ``Modern problems of
theoretical mathematics''.

\appendix
\section{Proof of Lemma 1}\label{appendixA}

Note that it does not restrict generality to take
$\sigma^{E}_{2\ell}=\sigma_{2\ell}$, as this can always
be accompanied by an appropriate similarity transform.
Our problem at hand
of extending a symplectic form is then
equivalent to the following problem: Suppose we are given
column vectors $e_1,\cdots, e_n$ and $f_1,\cdots, f_n$
from $\rr^{2(n+\ell)}$
that satisfy
\begin{eqnarray}
    e_j^T \sigma_{2(n+\ell)} e_k &=& 0,\\
    f_j^T \sigma_{2(n+\ell)} f_k &=& 0,\\
    e_j^T \sigma_{2(n+\ell)} f_k &=& \delta_{j,k},
\end{eqnarray}
for $j,k=1,\cdots, n$. The procedure continues by
identifying vectors $e_{n+1}$ and $f_{n+1}$ such that
$e_{n+1}^T \sigma_{2(n+\ell)} f_{n+1}=1$ and
\begin{equation}
    e_{n+1}^T \sigma_{2(n+\ell)} w=f_{n+1}^T \sigma_{2(n+\ell)} w=
    0
\end{equation}
for all
\begin{equation}
    w\in W_n:=\text{span}(e_1,\cdots, e_n, f_1,\cdots, f_n).
\end{equation}
Now define
\begin{equation}
    W_n^\perp = \{w: w^T \sigma_{2(n+\ell)} v =0\, \forall
    v\in W_n\}.
\end{equation}
It is now not difficult to see that $W_n\cap W_n^\perp=\{0\}$
and $\rr^{2(n+\ell)}=W_n\oplus W_n^\perp$: Suppose
that the vector $v$ has
$v^T  \sigma_{2(n+\ell)} e_j=:\alpha_j$
and
$v^T  \sigma_{2(n+\ell)} f_j=:\beta_j$ for $j=1,\cdots, n$.
Then
\begin{equation}
    v=\left[ \sum_{j=1}^n \left(
    -\alpha_j f_j + \beta_j e_j
    \right)\right]+
     \left[v+ \sum_{j=1}^n \left(
    \alpha_j f_j - \beta_j e_j\right)
    \right],
\end{equation}
where the first term is element of $W_n$ and the
second of $W_n^\perp$.
Following a symplectic Gram-Schmidt procedure,
the symplectic basis can hence
be completed, which is equivalent
to extending the matrices $s_1$ and $s_2$ to a symplectic
\begin{equation}
    S=\left[
    \begin{array}{cc}
    s_1 & s_2 \\
    s_3 & s_4
    \end{array}
    \right]\in Sp(2(n+\ell),\rr).
\end{equation}

\section{Derivation of Eq.~(\ref{prob2})} \label{AppendixB}

Here we show that Eq.~(\ref{prob2}) admits solution for
$s_2^\prime$ as in Eq.~(\ref{defs2prime}). In fact, assuming
$\sigma_{4n}^E = \sigma_{2n} \oplus \sigma_{2n}$ with
$\sigma_{2n}$ as in Eq.~(\ref{NSYMP}), one has
\begin{eqnarray}\label{Nall}
s_2^\prime \, \sigma^E_{4n}  \,(s_2^\prime)^T -\Sigma^\prime &=&
   \left[
    \begin{array}{c|c}
     K^{-1} & O^T A
    \end{array}
    \right]
     \left[
    \begin{array}{c|c}
    \sigma_{2n} & 0 \\ \hline
     0 & \sigma_{2n}
    \end{array}
    \right]
        \left[
    \begin{array}{c}
    K^{-T} \\ \hline   A^T O
    \end{array}
    \right]  - \Sigma^\prime \nonumber \\
    &=&
   K^{-1} \sigma_{2n} K^{-T}  +
     O^{T} A \, \sigma_{2n} \,  A^T O - \Sigma^\prime \nonumber \\
     &=& K^{-1} \, \left( K \Sigma^\prime K^T  + B  \right)\, K^{-T}  +
     O^{T} A \, \sigma_{2n} \,  A^T O - \Sigma^\prime \nonumber \\
      &=& K^{-1} \, B   K^{-T}  +
     O^{T} A \, \sigma_{2n} \,  A^T O \nonumber\\
     &=& O \left( M^{1/2} B M^{1/2} + A \, \sigma_{2n}\,
     A^T \right) O^T\,,
\end{eqnarray}
where we used  Eq.~(\ref{NIDENTITY}) to write $\sigma_{2n} =
K \Sigma^\prime K^T  + B$, with $B$ being the $2n \times 2n$ matrix
   \begin{eqnarray} B&:=&
    \left[
    \begin{array}{c|c}
    {0} & \, \begin{array}{c|c}
  \;\;\;\;\;0 \;\;\;\;\;& \;\;\;\;\;0\;\;\;\;\; \\  \hline
    0 & \openone_{n-r/2}
    \end{array}
        \\ \hline
     \begin{array}{c|c}
    \;\;\;\;\;0 \;\;\;\;\;& \;\;\;\;\;0\;\;\;\;\; \\ \hline
    0 & -\openone_{n-r/2}
    \end{array}
    & \, {0}
    \end{array}
    \right]\,. \end{eqnarray}
The identity~(\ref{prob2})
finally follows by noticing that  the last term
in Eq.~(\ref{Nall}) cancels
 since $M^{1/2} B = BM^{1/2} = B$
 and $A \, \sigma_{2n} \, A^T = -B$.

\section{Properties of the environmental states} \label{AppendixC}

In this appendix we first give an explicit derivation of
Eq.~(\ref{NNequation}). Then we  analyze in details the property
of the state $\hat{\rho}_E$ associated with the covariance matrix
$\gamma_E$ defined be the Eqs.~(\ref{Nalpha}) and (\ref{Ndelta}).
Replacing Eq.~(\ref{defs2prime}) into Eq.~(\ref{probb1}), we get
    \begin{eqnarray}
    \openone_{2n} =s_2^\prime \,  \gamma_E  \,
    (s_2^\prime)^T  &=&
     \left[
    \begin{array}{c|c}
    K^{-1} & O^T A
    \end{array}
    \right] \,  \left[
    \begin{array}{c|c}
    \alpha & \, \delta \\ \hline
    \delta^T & \beta
     \end{array}
    \right] \,  \left[
    \begin{array}{c}
    K^{-T} \\ \hline   A^T O
    \end{array}
    \right] \nonumber \\
&=& K^{-1} \,
\alpha \, K^{-T} + O^T \, A \, \delta^T \,K^{-T} +  K^{-1} \,
\delta \,A^T \,O + O^T \, A \, \beta \,A^T \, O \nonumber \\
&=& O^T \left( M^{1/2} \, \alpha \, M^{1/2} +  A \, \delta^T
\,M^{1/2} +  M^{1/2} \,  \delta \,A^T  +  A \, \beta \,A^T \right)
O \,, \nonumber
    \end{eqnarray}
 which leads to
 \begin{eqnarray}
 M^{-1} = \alpha +   M^{-1/2} A \, \delta^T  +   \delta \,A^T  M^{-1/2}  + M^{-1/2} \,  A \, \beta \,A^T M^{-1/2}  \,,
 \end{eqnarray}
 and hence to Eq.~(\ref{NNequation}) by the fact $M^{-1/2} A = A^T   M^{-1/2}   = A =A^T$.
Such an equation admits the solution given in Eqs.~(\ref{Nalpha})
and (\ref{Ndelta}). Explicitly this corresponds to the $4n\times 4
n$ covariance matrix $\gamma_E$ of the form
  \begin{eqnarray}
  \left[
  \begin{array}{c|c|c|c}
              \begin{array}{c|c}
   \mu^{-1} &\,\,\,\,\,\,0\,\,\,\,\,\, \\ \hline
  \,\,\,\,\,0\,\,\,\,\,\,\, & \xi\openone
    \end{array}
& 0 & 0 & \begin{array}{c|c}
    f(\mu^{-1})  &0 \\ \hline
    0 &  f(\xi \openone)
    \end{array} \\ \hline
    0&
           \begin{array}{c|c}
     \mu^{-1} &\,\,\,\,\,\,0\,\,\,\,\,\, \\ \hline
  \,\,\,\,\,0\,\,\,\,\,\,\, & \xi\openone
    \end{array}  &
   \begin{array}{c|c}
    f(\mu^{-1})  &\;\,\,\,\,\;0\;\;\,\,\,\, \\ \hline
    \,\,\,\,\;\;0\,\,\,\,\;\;&  f(\xi \openone)
    \end{array} & 0
    \\ \hline
       0  &
        \begin{array}{c|c}
    f(\mu^{-1})  &\;\,\,\,\,\;0\;\;\,\,\,\, \\ \hline
   \;\,\,\,\,\;0\;\;\,\,\,\,  &  f(\xi \openone)
    \end{array} &
        \begin{array}{c|c}
    \mu^{-1} &\,\,\,\,\,\,0\,\,\,\,\,\, \\ \hline
  \,\,\,\,\,0\,\,\,\,\,\,\, & \xi\openone
      \end{array} & 0 \\\hline
     \begin{array}{c|c}
    f(\mu^{-1})  &\,\,\,\,\,0\,\,\,\,\, \\ \hline
    \,\,\,\,\,0\,\,\,\,\, &  f(\xi \openone)
    \end{array} & 0 & 0 & \begin{array}{c|c}
    \mu^{-1} &\,\,\,\,\,\,0\,\,\,\,\,\, \\ \hline
  \,\,\,\,\,0\,\,\,\,\,\,\, & \xi\openone
    \end{array}
    \end{array}
    \right] \nonumber
    \end{eqnarray}
where for easy of notation $\openone := \openone_{n -r/2}$.
By looking at the structure of this covariance matrix, one
realizes that it is composed by two independent sets formed by $r$
and $2n-r$ modes, respectively. The first set describes $r/2$
thermal states characterized by the matrices $\mu^{-1}$ which have
been purified  adding further $r/2$ modes. The second  set
instead  describes a collection of $2(n-r/2)= 2n -r$ modes
prepared in a pure state formed by $n-r/2$ independent pairs
of modes which are entangled. By reorganizing its rows and
columns
this can be cast into the simpler form
\begin{eqnarray}\label{defAprime}
    \gamma_E  &=& \left[
    \begin{array}{c|c}
     \begin{array}{c|c}
    \bar{\mu}^{-1} & f( \bar{\mu}^{-1}) \\  \hline
     f( \bar{\mu}^{-1})  &  \bar{\mu}^{-1}
    \end{array}
     & 0 \\  \hline
    0 &  \begin{array}{c|c}
  {\xi \openone_{2n-r}} & f({\xi \openone}_{2n-r})  \\  \hline
    f({\xi \openone}_{2n-r}) &  {\xi \openone}_{2n-r}
    \end{array}
    \end{array}
    \right] \begin{array}{l}
    \} \, r \\
    \} \, r \\
     \} \, 2n -r \\
    \} \, 2 n-r \,,\\
    \end{array} \label{nuovaversdigamma}
    \end{eqnarray}
    where we used $\bar{\mu}$ to indicate the  $r\times r$
matrix $\bar{\mu}=\mu \oplus \mu$.

\subsection{Solution for $\ell_{\text{pure}}
= 2n -r^\prime/2$ environmental modes} \label{app:sols}

Defining $r^\prime$ as in Eq.~(\ref{defrprime}) we choose the
environmental commutation matrix to be $\sigma_{2\ell}^E =
\sigma_{2n} \oplus \sigma_{2n -r^\prime}$ with $\sigma_{2n}$ and
$\sigma_{2n-r^\prime}$ as in Eq.~(\ref{NSYMP}). A unitary dilation
with $\ell_{\text{pure}} =2n -r^\prime/2$ environmental modes in a
pure state is obtained by having $s_2 = Y^{1/2} s_2^\prime$ with
$s_2^\prime$   as in Eq.~(\ref{defs2prime}). In this case,
however,
$A$ is a rectangular matrix $2 n\times 2 (n-r^\prime/2)$ of the
form
 \begin{eqnarray}\label{defAnew} A &=& \left[
    \begin{array}{c|c}
    {0} &  \begin{array}{c|c}
     \;\;\;\; 0\;\;\;\; & 0 \\ \hline
    0 &  0
        \\ \hline
        0 & \openone_{n-r/2}
    \end{array}
        \\ \hline
\begin{array}{c|c}
       \;\;\;\; 0\;\;\;\; & 0 \\ \hline
    0 &  0
        \\ \hline
        0 & \openone_{n-r/2}
    \end{array}
    & {0}
    \end{array}
    \right] \begin{array}{l}
    \} \, r^\prime/2 \\
    \} \, (r-r^\prime)/2 \\
    \} \, n-r/2 \\
    \} \, r^\prime/2 \\
    \} \, (r-r^\prime)/2 \\
    \} \, n-r/2 .\\
    \end{array}
    \end{eqnarray}
    Similarly, the covariance matrix $\gamma_E$ can be
still expressed as in  Eq.~(\ref{gammaedef}). In this case, yet,
$\alpha$ is a $2n\times 2n$ matrix of block form
  \begin{eqnarray}
    \alpha =
     \left[
    \begin{array}{c|c}
        \begin{array}{c|c|c}
  \openone_{r'/2} &0&0 \\ \hline
  0&  \mu_o^{-1} & 0 \\ \hline
  0&   0 &  \xi \openone_{n-r/2}
    \end{array}
& 0 \\ \hline
    0 &          \begin{array}{c|c|c}
  \openone_{r'/2} &0&0 \\ \hline
  0&  \mu_o^{-1} & 0 \\ \hline
  0&   0 &  \xi \openone_{n-r/2}
    \end{array}
    \end{array}
    \right]
    \begin{array}{l}
    \} \, r^\prime /2 \\
    \} \, (r-r^\prime)/2 \\
    \} \, n-r/2\\
    \} \, r^\prime/2 \\
     \} \, (r -r^\prime)/2 \\
    \} \, n-r/2 , \\
    \end{array}\label{Nalphaii}
    \end{eqnarray}
where $\xi =5/4$ and $\mu_o$ is the $(r-r^\prime)/2\times
(r-r^\prime)/2$ diagonal matrix formed by the elements of $\mu$
which are strictly smaller than $1$. $\beta$ is the $(2n
-r^\prime) \times (2n-r^\prime)$ matrix
      \begin{eqnarray}
   \beta=
     \left[
    \begin{array}{c|c}
        \begin{array}{c|c}
  \;\;\; \mu_o^{-1} \;\;\;& 0 \\ \hline
    0 &  \xi \openone_{n-r/2}
    \end{array}
& 0 \\ \hline
    0 &          \begin{array}{c|c}
  \;\;\;\mu_o^{-1} \;\;\;& 0 \\ \hline
    0 &  \xi \openone_{n-r/2}
    \end{array}
    \end{array}
    \right]
    \begin{array}{l}
     \} \, (r-r^\prime)/2 \\
    \} \, n-r/2\\
      \} \, (r -r^\prime)/2 \\
    \} \, n-r/2 , \\
    \end{array}\label{Nbetaii}
    \end{eqnarray}
    and
 \begin{eqnarray}
    \delta = \left[
    \begin{array}{c|c}
        0
& \begin{array}{c|c}
0 & 0 \\ \hline
 \;\;\;f(\mu_o^{-1})\;\;\; & 0 \\ \hline
    0 &  f(\xi \openone_{n-r/2})
    \end{array} \\ \hline
     \begin{array}{c|c}
     0&0 \\ \hline
   \;\;\;  f(\mu_o^{-1})  \;\;\;& 0 \\ \hline
    0 &   f(\xi \openone_{n-r/2})
    \end{array}  &  0
    \end{array}
    \right]  \begin{array}{l}
    \} \, r^\prime /2 \\
    \} \, (r-r^\prime)/2 \\
    \} \, n-r/2\\
    \} \, r^\prime/2 \\
     \} \, (r -r^\prime)/2 \\
    \} \, n-r/2 , \\
    \end{array} \label{Ndeltaii}
    \end{eqnarray}
with $f$ as in Eq.~(\ref{Ndelta}).

By looking at the structure of this covariance matrix, one
realizes that it is composed by three independent pieces. The
first one describes a collection of $r^\prime/2$ vacuum states.
The second one, in turn,
describes $(r-r^\prime)/2$ thermal states
characterized by the matrices $\mu_o^{-1}$ which have
been purified by adding further
$(r-r^\prime)/2$ modes. The third
one, finally,  reflects a collection of $2(n-r/2)= 2n -r$ modes
prepared in a pure state formed by $n-r/2$ independent pairs
of modes which are entangled.

\subsection{Solution for $\ell = 2n -r/2$ not necessarily pure environmental modes}
\label{app:sols2}

In this subsection, we present the alternative
derivation of a dilation that does not necessarily involve
an environment prepared in a pure state.
Choosing the commutation matrix $\sigma_{2\ell}^E = \sigma_{2n}
\oplus \sigma_{2n -r}$ with $\sigma_{2n}$ and $\sigma_{2n-r}$ as
in Eq.~(\ref{NSYMP}), the matrix $s_2^\prime$   can be still
expressed as in Eq.~(\ref{defs2prime}). In this case, however,
$A$ is a rectangular matrix $2n\times (2n-r)$ of the form
 \begin{eqnarray}\label{defAnew1} A &=& \left[
    \begin{array}{c|c}
    {0} & \begin{array}{c}
  0 \\  \hline
 \openone_{n-r/2}
    \end{array}
        \\ \hline
     \begin{array}{c}
    0 \\ \hline
     \openone_{n-r/2}
    \end{array}
    & {0}
    \end{array}
    \right] \begin{array}{l}
    \} \, r/2 \\
    \} \, n-r/2\\
     \} \, r/2 \\
    \} \, n-r/2 ,\\
    \end{array}.
    \end{eqnarray}
Similarly,
$\gamma_E$ has the
block form~(\ref{gammaedef}), where  $\alpha$ is
still the $2n \times 2n$ matrix of Eq.~(\ref{Nalpha}), while
$\beta$ and $\delta$ are, respectively, the
following $(2n -r)\times (2n-r)$ and
 $2n \times (2n-r)$ real matrices:
     \begin{eqnarray}
    \beta =
     \left[
    \begin{array}{c|c}
        \xi \openone_{n-r/2}
& 0 \\ \hline
    0 &          \xi \openone_{n-r/2}     \end{array}
    \right]
    \begin{array}{l}
     \} \, n-r/2\\
     \} \, n-r/2 , \\
    \end{array}\label{Nbeta}
    \end{eqnarray}
  \begin{eqnarray}
    \delta = \left[
    \begin{array}{c|c}
      0 & 0 \\ \hline
    0 &  f(\xi \openone_{n-r/2})
        \\ \hline
        0 & 0 \\ \hline
    f(\xi \openone_{n-r/2}) & 0
    \end{array}
    \right]  \begin{array}{l}
    \} \, r/2 \\
    \} \, n-r/2\\
     \} \, r/2 \\
    \} \, n-r/2 , \\
    \end{array}
    \end{eqnarray}
    with $\xi$ and $f$ as in Eq.~(\ref{Ndelta}). That is,
  \begin{eqnarray} \label{vgammae}
\gamma_E =\left[
  \begin{array}{c|c|c}
              \begin{array}{c|c}
    \mu^{-1} &\,\,\,\,\,\,\,0\,\,\,\,\,\,\,\\ \hline
 \,\,\,\,\,\,\,0\,\,\,\,\,\,\, & \xi\openone
    \end{array}
&  0  & \begin{array}{c|c}
    \,\,\,\,\,\,\,0\,\,\,\,\,\,\, &\,\,\,\,\,\,\,0\,\,\,\,\,\,\, \\ \hline
   0&  f(\xi \openone)
    \end{array} \\ \hline
    0&
           \begin{array}{c|c}
    \mu^{-1} &\,\,\,\,\,\,\,0\,\,\,\,\,\,\, \\ \hline
  \,\,\,\,\,\,\,0\,\,\,\,\,\,\, & \xi\openone
    \end{array}  &
    \begin{array}{c|c}
   \,\,\,\,\,\,\,0\,\,\,\,\,\,\, &\,\,\,\,\,\,\,0\,\,\,\,\,\,\, \\ \hline
  f(\xi \openone) & 0
    \end{array} \\ \hline \hline
        \begin{array}{c|c}
    \,\,\,\,\,\,\,0\,\,\,\,\,\,\,&\,\,\,\,\,\,\,0\,\,\,\,\,\,\, \\ \hline
0 &  f(\xi \openone)
    \end{array}  &
        \begin{array}{c|c}
   \,\,\,\,\,\,\,0\,\,\,\,\,\,\, & f(\xi \openone)   \\ \hline
0 &\,\,\,\,\,\,\,0\,\,\,\,\,\,\,   \end{array} &
        \begin{array}{c|c}
    \xi \openone &\,\,\,\,\,\,\,0\,\,\,\,\,\,\, \\ \hline
 \,\,\,\,\,\,\,0\,\,\,\,\,\,\,& \xi\openone
    \end{array}
    \end{array}
    \right] \begin{array}{l}
    \} \, r/2 \\
    \} \, n-r/2\\
     \} \, r/2 \\
    \} \, n-r/2 \\
    \} \, n-r/2 \\
\}\, n-r/2 \,,\\
    \end{array}
    \end{eqnarray}
with $\openone= \openone_{n-r/2}$.
This covariance matrix now consists of two
independent parts: The first one describes
a collection of $r/2$ thermal states described by the matrices
$\mu^{-1}$. The second instead reflects a collection of
$2(n-r/2)= 2n -r$ modes prepared in a pure state formed by $n-r/2$
independent couples of modes which are entangled.  The covariance
matrix given in Theorem 1 can be recovered from the one given above
by adding $r$ modes to purify the
thermal states $\mu^{-1}$.

\section{Equivalent unitary dilations}
\label{sec:equivalent}

Let
\begin{equation}\label{sm}
    S = \left[
    \begin{array}{cc}
        s_1 &s_2\\
        s_3 &s_4
    \end{array}\right]
\end{equation}
and $\gamma_E$ define a unitary dilation for a bosonic Gaussian
channel $\Phi$ characterized by matrices $X$ and $Y$. Then a full
class of unitary dilations
\begin{equation}\label{spm}
    S' = \left[
    \begin{array}{cc}
        s_1' &s_2'\\\
        s_3' &s_4'
    \end{array}\right]
\end{equation}
can be obtained by taking $\gamma_E^\prime =
V \gamma_E V^T$ and \bea \label{Nequi} s_1^\prime = s_1\,, \quad
s_2'=s_2 V\, \quad  \, s_3'= W s_3\,,  \quad  \, s_4'= W s_4 V\,,
 \eea
with $V \in Sp(2 \ell,\rr) $ and $W\in Sp(2n,\rr)$ being symplectic
transformations of $\ell$ and $n$ modes respectively. With this
choice in fact $\gamma_E'$ is still a covariance matrix while the
conditions~(\ref{NSN}) and (\ref{dil}) are automatically satisfied.
From a physical point of view the symplectic transformations $V$ and
$W$ correspond to unitary local operations applied to the
environmental input and output states, respectively, by virtue of
the metaplectic representation. Consequently, the weak complementary
channels $\tilde{\Phi}$ and $\tilde{\Phi}'$ associated with these
two representations are unitarily equivalent and  the
weak-degradability properties one can determine for $\Phi$ will be
the same when studied for $\Phi'$.

Conversely, let us suppose to have two unitary dilations of $\Phi$,
realized with $\ell=n$ environmental modes and characterized by the
symplectic matrices $S$ and $S^\prime$ as in Eq.\ (\ref{sm}) and
(\ref{spm}), respectively, with $s_i$ and $s_i'$ being $2n\times 2n$
square matrices. Then it is possible  to show that they must be
related as in Eq.~(\ref{Nequi}) under the  hypothesis that  $s_2$
and $s_3$ are non-singular. First of all, since Eq.~(\ref{dil}) must
be satisfied for all the input covariance matrices  $\gamma$, we
have $s_1=X^T=s_1^\prime$. Define then  $V = s_2^{-1} s_2'$ and $W =
s_3' s_3^{-1} $. By using the first of Eq.~(\ref{NSN}) and
exploiting the non-singularity of $s_2$ one has
 \bea
 s_2\, V \,\sigma_{2\ell}^E \,V^T \,s_2^T = s_2 \, \sigma_{2n}\, s_2^T
 \quad \Longrightarrow \quad V \, \sigma_{2 n} \, V^T=\sigma_{2 n}\,,
  \eea
which implies that $V$ is a symplectic matrix
(we are assuming $\sigma_{2\ell}^E = \sigma_{2n}$).
Moreover, from the second
condition in Eqs.\ (\ref{NSN}) for $S$ and $S'$, we obtain
 \bea
 s_2 \sigma s_4^T W^T = s_2 V \sigma s_4'^T \,, \qquad  \Longrightarrow \qquad
  s_4'= W s_4 V\,,
 \eea
because $s_2$ is non-singular and $V$ is symplectic. By considering
the third condition (\ref{NSN}) one then has
 \bea
W ( s_3 \sigma_{2n} s_3^T + s_4 \sigma_{2n} s_4^T) W^T = W
\sigma_{2n} W^T = \sigma_{2n}
 \eea
which prove that $W$ is a symplectic. Finally, let us observe that
the proof above does not use the non-singularity of $s_3$. Indeed,
one can relax this hypothesis and assume more simply that there
exists a $W$ such that $s_3'=  W s_3$; from
Eqs.\ (\ref{NSN}) $W$ has
to still be a symplectic matrix but $s_3$ and $s_3'$ may be
singular.

As an application of these equivalent unitary dilation results, we
can find an alternative canonical form to the one in Sec.
\ref{sec:canonical} with the same $s_1$ and $s_4$ but with $s_2$
and $s_3$ of the following anti-diagonal block form
\begin{equation}
    s_{j} = \left[%
 \begin{array}{cc}  0 & F_j \\ G_j & 0 \\
\end{array}%
\right]
\end{equation}
where, for $j=2,3$,  $F_j$, $G_j$ are $n \times n$ real
matrices. Imposing Eqs.\ (\ref{NSN}), one obtains the following
relations
 \bea
J^T- F_2 G_2^T &=& \openone_n
\ \, \ , \ \, \ J'^T- F_3 G_3^T = \openone_n  , \\
F_2 - F_3^T &=& 0  \ \, \ , \ \, \ J G_3^T - G_2 J'^T = 0
\nonumber \, ,
 \eea
the solution of which provides
the following unitary dilation,
 \bea
S=\left[%
\begin{array}{cc|cc}
  \openone_n & 0 & 0 & -(\openone_n-J^T)G_2^{-T} \\
  0 & J & G_2 & 0 \\ \hline
  0 & -G_2^{-1} (\openone_n-J) & \openone_n & 0   \\
  G_2^T & 0 & 0 & G_2^{T}J^T G_2^{-T} \\
\end{array}%
\right] \, ,
 \eea
where again $G_2$ is an arbitrary (non-singular) matrix and the
eigenvalues of $J$ are assumed to be different from $1$.
This solution is unitarily equivalent to the one in Eq.\
(\ref{Sprecanonical}) by applying
$V=-\sigma_{2n}$ and
\begin{equation}
W = \left[%
\begin{array}{cc}
  0 & G_2^{-1} J^{-1} G_2 \\
  -G_2^T J^T G_2^{-T} & 0 \\
\end{array}%
\right]
\end{equation}
as above.

\section{The ideal-like quantum channel}
\label{ideal}

Here we consider a quantum channel with $X=\openone_{2n}$ but $Y
\geqslant 0$ with rank less than $2n$, which can be described in
terms of only $n$ additional (environmental) modes. We call it
ideal-like quantum channel. Accordingly, the canonical unitary
transformation $\hat{U}$ of Eq.~(\ref{eq1}) will be uniquely
determined by a $4n \times 4n$ real matrix $S\in Sp(4n,\rr)$ of
block form in Eq.~(\ref{NmatrixS}), where $s_i$ are $2n \times 2n$
real matrices. Particularly, $s_1=s_4=\openone_{2n}$,
\begin{equation}
s_{3} = \left[%
 \begin{array}{cc}  F_3 & {0} \\  {0} & G_3
\end{array}%
\right],\qquad  s_{2} = \left[%
 \begin{array}{cc}  -G_3^T & {0} \\ {0} & -F_3^T
\end{array}%
\right],
\end{equation}
with $F_3$ and $G_3$ being $n \times n$ real matrices such
that $F_3 G_3^T=G_3^T F_3=0$, in order to satisfy the symplectic
conditions in Eqs. (\ref{NSN}). Taking advantage of the freedom in
the choice of the unitary dilation shown in Appendix
\ref{sec:equivalent}, the matrix $S$ can be
put in the form of Eq.~(\ref{NmatrixS}) in which
$s_1'=s_4'=\openone_{2n}$,
\begin{equation}
s_{2}' = \left[%
 \begin{array}{cc}  0 & 0 \\  0 & \openone_n
\end{array}%
\right],\qquad
s_{3}' = \left[%
 \begin{array}{cc} -\openone_n  & \;\;0\;\; \\ 0 & 0
\end{array}%
\right]\;,
\end{equation}
 where $F_3$ is assumed non-singular. In this respect, one
uses $V, W \in Sp(2n,\rr)$ (of App. \ref{sec:equivalent})
of the following form
\begin{equation}
V=\left[%
\begin{array}{cc}
  -F_3 & 0 \\
  0 & -F_3^{-T} \\
\end{array}%
\right]\;,
\end{equation}
 and $W=V^{-1}$. Similarly, one can proceed, if $G_3$ is
non-singular, and obtains a similar structure for $S$ as above. As
concerns the weak-degradability properties, if one assumes the
initial environmental input state as $\gamma_E=\textrm{diag}(2N+1,
2M+1, 2N+1, 2M+1)$, the eigenvalues of
$\tilde{Y} - \tilde{X}^T
X^{-T} (Y+i \sigma) X^{-1} \tilde{X} + i \sigma$ are
$\{2M,2(M+1),2N,2(N+1)\}$, which are always positive for any $N
\geqslant 0$ and $M \geqslant 0$; hence, this channel with $\gamma_E$ as
above is always weakly degradable.

Finally, one may consider another ideal-like channel with
$X=\openone_{2n}$ and $Y=[(1-\sigma_3)/4]^{\otimes n}$, i.e.
$\Phi_{X,Y}=\bigotimes_{i=1}^{n} ({B_1})_i$, where the
single-mode $B_1$ channel is
defined in Ref.\ \cite{CGH} as $X=\openone_2$ and
$Y=(1-\sigma_3)/4$. Trivially, this multi-mode channel is always
WD (like $B_1$) and is able to transfer a quantum state without
decoherence with the maximum quantum capacity (like
for the single-mode case \cite{CGH}).

\end{document}